# Reversing Hydrogen-Related Loss in α-Ta Thin Films for Quantum Device Fabrication

D. P. Lozano[1], M. Mongillo[1], B. Raes[1], Y. Canvel[1], S. Massar[1], A. M. Vadiraj[1], Ts. Ivanov[1], R. Acharya[1], J. Van Damme[1,2], J. Van de Vondel[3], D. Wan[1], A. Potočnik[1], and K. De Greve[1,2]

[1]Imec, Kapeldreef 75, Leuven, B-3001, Belgium,

[2]Department of Electrical Engineering (ESAT), KU Leuven, Leuven, B-3001, Belgium

[3]Departmentof Physics and Astronomy, KU Leuven, Leuven, B-3001, Belgium

E-mail: daniel.perezlozano@imec.be

E-mail: anton.potocnik@imec.be

## Abstract

α-Tantalum (α-Ta) is an emerging material for superconducting qubit fabrication due to the low microwave loss of its stable native oxide. However, hydrogen absorption during fabrication, particularly when removing the native oxide, can degrade performance by increasing microwave loss. In this work, we demonstrate that hydrogen can enter α-Ta thin films when exposed to 10 vol% hydrofluoric acid for 3 minutes or longer, leading to an increase in power-independent ohmic loss in high-Q resonators at millikelvin temperatures. Reduced resonator performance is likely caused by the formation of non-superconducting tantalum hydride ($TaH_x$) precipitates. We further show that annealing at 500°C in ultra-high vacuum ($10^{-8}$ Torr) for one hour fully removes hydrogen and restores the resonators' intrinsic quality factors to ~4 million at the single-photon level. These findings identify a previously unreported loss mechanism in α-Ta and offer a pathway to reverse hydrogen-induced degradation in quantum devices based on Ta and, by extension also Nb, enabling more robust fabrication processes for superconducting qubits.

## Introduction

Superconducting qubits have garnered significant attention in quantum information technologies due to their scalability and high-gate fidelity.[1] Despite considerable advancements in qubit performance over the past two decades, further improvements in coherence times and fidelity are essential for realizing practical, large-scale quantum computers based on superconducting qubits.[2] Among the various sources of decoherence, fabrication processes play a critical role, as they can introduce defects and impurities that contribute to microwave loss that in turn limit coherence: examples of such defects include two-level systems and excess quasiparticles.[3–5] To achieve high coherence while maintaining compatibility with industrial-scale methods, qubit fabrication can leverage processing techniques borrowed from the complementary metal-oxide-semiconductor (CMOS) industry.[6]

α-Tantalum (α-Ta) has recently emerged as a promising material for superconducting qubits due to its favorable morphological and superconducting properties, including low microwave loss of its stable native oxide.[7–13] These characteristics make α-Ta a strong alternative to aluminum (Al) and niobium (Nb),[14] which are widely used in superconducting qubit fabrication.[4] In particular, Ta's native oxide exhibits a more stable and well-coordinated atomic structure with lower oxygen deficiency compared to Nb oxides, likely resulting in lower magnetic moments density.[15] Despite these advantages, α-Ta—like Nb—has a high hydrogen solubility due to its large interstitial sites in the body-centered-cubic (bcc) crystal structure.[16–18]

The presence of diluted hydrogen in Nb metal has been linked to a reduction of high Q-factors in superconducting radio frequency cavities[19–21] and, more recently, in planar superconducting Nb resonators[22] fabricated using processes similar to those of superconducting qubits.[23,24] Hydrogen absorption in Nb—and likely also in α-Ta films—can occur during various fabrication steps, particularly when the protective native oxide layer, which serves as a hydrogen diffusion barrier,[21,22,25] is removed. Hydrogen incorporation can take place during oxide cleaning with fluorine-based etchants, such as hydrofluoric acid (HF)[9,26] or less aggressive buffered oxide etchant (BOE),[10,11,27] both commonly used to remove surface oxides and passivate silicon-air interface. Additionally, hydrogen can enter the metal during acidic or hydrogen-rich etching processes (e.g. chlorine- of fluorine-based wet[7] or dry[8,9,14] etching), either through the exposed metal surface during etching or at the sidewalls after etching, where the metal remains unprotected during venting to atmosphere.

While the presence of hydrogen, in the form of niobium hydrides ($NbH_x$) precipitates,[21,23] and its detrimental impact on microwave loss in high-Q resonators[22] have been established for Nb-based devices, no evidence of tantalum hydrides ($TaH_x$) formation or its potential effect on high-Q resonators and qubits has been reported or systematically investigated for α-Ta thin films.

In this study, we investigate the impact of hydrogen absorption in α-Ta thin films, with a particular focus on its effect on microwave loss relevant to superconducting qubit performance. We show that hydrogen can diffuse into α-Ta in an acidic environment, increasing power-independent ohmic loss and suppressing resonance features in superconducting resonators at millikelvin temperatures. However, we also demonstrate that annealing these films in ultra-high vacuum (UHV) at 500 °C for one hour removes hydrogen and fully restores the resonators' quality factors. These findings enable the use of diluted HF for faster and more aggressive surface cleaning and oxide removal, while also providing a technique to reverse the detrimental effects of hydrogen absorption in Ta-based quantum devices.

## Methodology

Hydrogen incorporation into α-Ta films is achieved through a hydrofluoric acid (HF) surface oxide removal process.[26] Samples with patterned coplanar waveguide resonators on α-Ta films and silicon wafers, fabricated using the high-temperature process described in Ref. 9, are submerged in 10 vol% diluted HF for varying durations: 0 min (reference), 1 min, 2 min, 3 min, 5 min, and 10 min. Immediately after the HF treatment, the samples are rinsed in deionized water until the water resistivity in the bath exceeded 12 MΩ.

Hydrogen absorption into the α-Ta films and further modifications due to the HF treatment are investigated thoroughly on patterned samples with the following material characterization techniques: scanning transmission electron microscopy (STEM), time-of-flight secondary ion mass spectrometry (ToF-SIMS), elastic recoil detection analysis (ERDA), atomic force microscopy (AFM) and X-ray photoelectron spectroscopy (XPS). The same set of samples is used for ToF-SIMS, AFM and XPS measurements and different sets are used for the STEM and ERDA measurements (see Methods section in Supporting Information). In addition to the morphological investigations, high-Q resonator measurements are performed at 10 mK in a dilution refrigerator designed for high-coherence superconducting qubit measurements (see Methods section in Supporting Information).[9]

## Results and discussion

### Morphology

STEM analysis reveals that the etching process initiates at the triple point—where the metal-air, substrate-air, and substrate-metal interfaces intersect (Figure 1**a**). At this location, a small gap forms once $SiO_x$ is removed by HF (after ~1 min),[26] exposing the Ta film and making it susceptible to etching. As the duration of HF exposure increases, the etched area gradually extends deeper into the film. The same behavior is observed on several samples, four of which are presented in Figure 1**a.** STEM micrographs suggest that the triple point serves as the primary pathway for hydrogen infiltration into the film. This is supported by the observation that other regions, specifically the top Ta surface and a large part of the sidewall remain resistant to etching for at least the first ~5 minutes of HF exposure. After this time, the HF starts inducing substantial surface modification, indicated by the AFM measurements (Figure 1**b,c**). After 10 minutes HF removes approximately ~50 nm of tantalum from the sidewall and ~40 nm from the top surface. The observed difference in etching behavior arises from the varying chemical reactivity of pure α-Ta and tantalum oxide. Experiments demonstrate that even a brief 1-minute exposure to the HF solution is sufficient to initiate etching of unprotected α-Ta with an etching rate of approximately 15 nm $min^{-1}$. In contrast, tantalum oxide exhibits significantly greater resistance, with an estimated etching rate of 0.4 nm $min^{-1}$, a value that aligns with previously reported findings in the literature[28] (see Supporting Information for details).

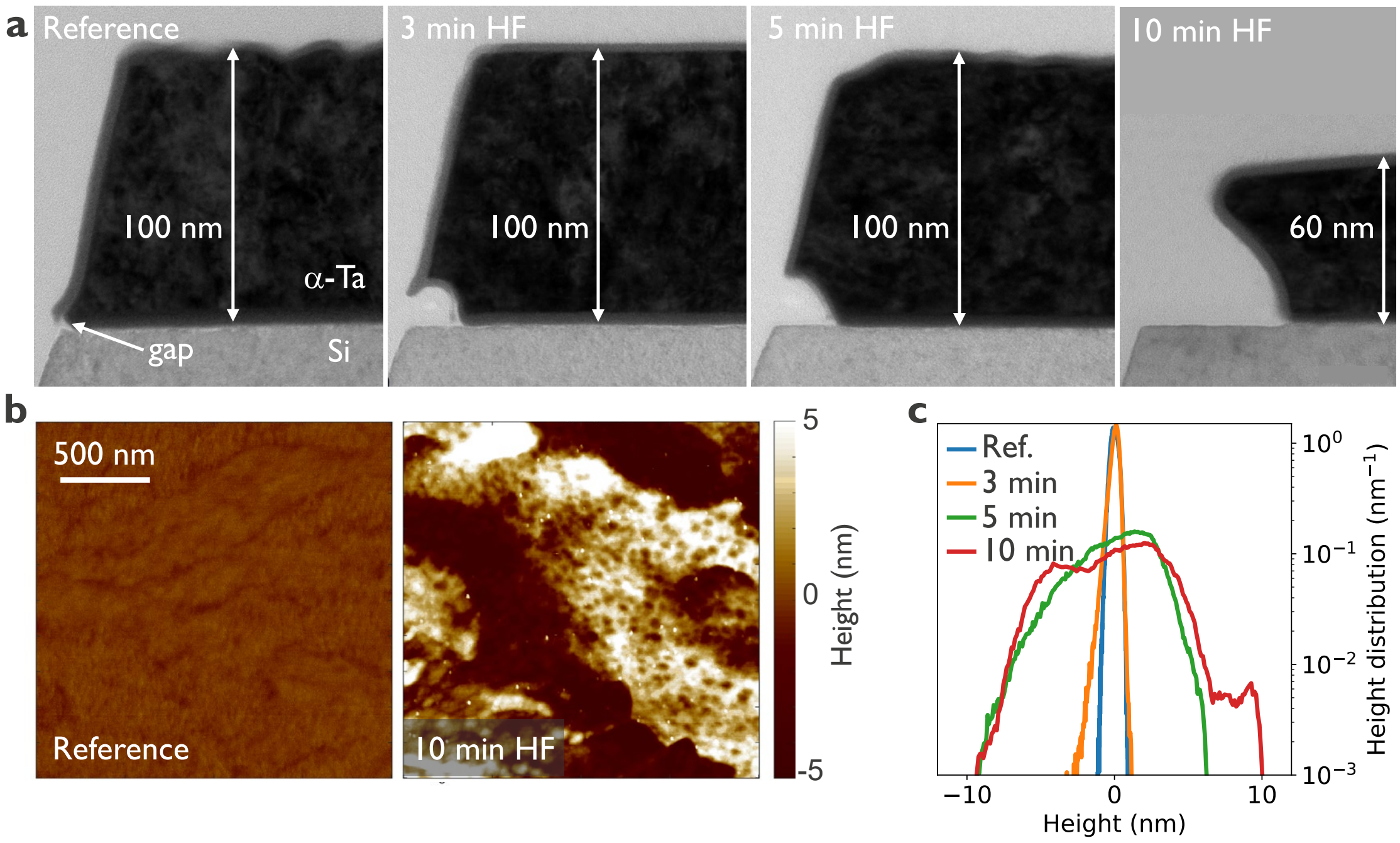

*Figure 1: Ta resonators exposed to diluted HF for different durations.* ***a*** *cross-STEM micrographs.* ***b*** *AFM topography of the Ta top surface for reference and 10-minute HF sample. AFM images of all other samples are shown in the Supporting Information.* ***c*** *Height probability density measured with AFM across the 2 μm x 2 μm area. Root-mean-square roughness ($R_q$) values for reference, 3 min, 5 min and 10 min samples are 0.28 nm, 0.38 nm, 2,65 nm and 3.30 nm, respectively.*

## Hydrogen content

The presence of hydrogen diluted in Ta films is first investigated with a ToF-SIMS measurements. To avoid hydrogen atoms rapidly discharging from the film into the high vacuum of the sample chamber as soon as the native oxide is sputtered away[25] an oxygen ($O_2$) sputtering beam is employed, which oxidizes the surface and prevents premature hydrogen desorption during the ToF-SIMS measurement. The normalized background level of hydrogen is ~$4 \cdot 10^{-4}$ in the untreated reference Ta sample (Figure 2**a**). Similar levels are found in samples exposed to HF for 2 min (Supporting Information). The amount of detected hydrogen starts increasing at 3 min and reaches levels of ~$1 \cdot 10^{-2}$ at 10 min exposure duration (Figure 2**b,c,d**). The detection of elevated hydrogen ($H^+$) levels throughout the film in the 3-minute sample, combined with the presence of an oxide layer on all other surfaces acting as a strong hydrogen diffusion barrier, suggests that a significant amount of hydrogen is absorbed through the triple points which relatively quickly spreads throughout the film. This is supported by the fact that the diffusion coefficient of hydrogen in Ta is approximately $10^{-6}$ $cm^2/s$ at room temperature,[17,18] allowing hydrogen to diffuse across hundreds of micrometers within minutes in patterned Ta film. In contrast to $H^+$, tantalum hydride species ($Ta_2H^-$, $TaH_5^-$, and $TaH_x^-$) are either absent or show levels comparable to the background.

To further confirm the presence of hydrogen in the HF treated α-Ta samples and identify possible tantalum hydride phases,[29] elastic recoil detection analysis (ERDA) is performed to quantitatively determine the atomic fraction of hydrogen in the metal. For the reference, 2 min, and 3 min samples, the hydrogen atomic fraction is comparable, ranging from 3% to 4% (Figure 2**e**), qualitatively consistent with the ToF-SIMS data. At these levels, hydrogen is randomly filling predominantly tetragonal interstitial sites in the bcc α-Ta at room temperature, resulting in a disordered solution of hydrogen in tantalum.[29] In the 5 min sample, however, the hydrogen atomic fraction increases to ~12%, placing it at the boundary between the α-Ta and α-Ta + ε phases,[29,30] suggesting that ε-tantalum hydride (ε-$Ta_2H$) might also be present. For the 10 min HF sample, the hydrogen atomic fraction reached 40%-45%, making δ-tantalum hydride (δ-$Ta_2H$) also a potential hydride phase present in the film.

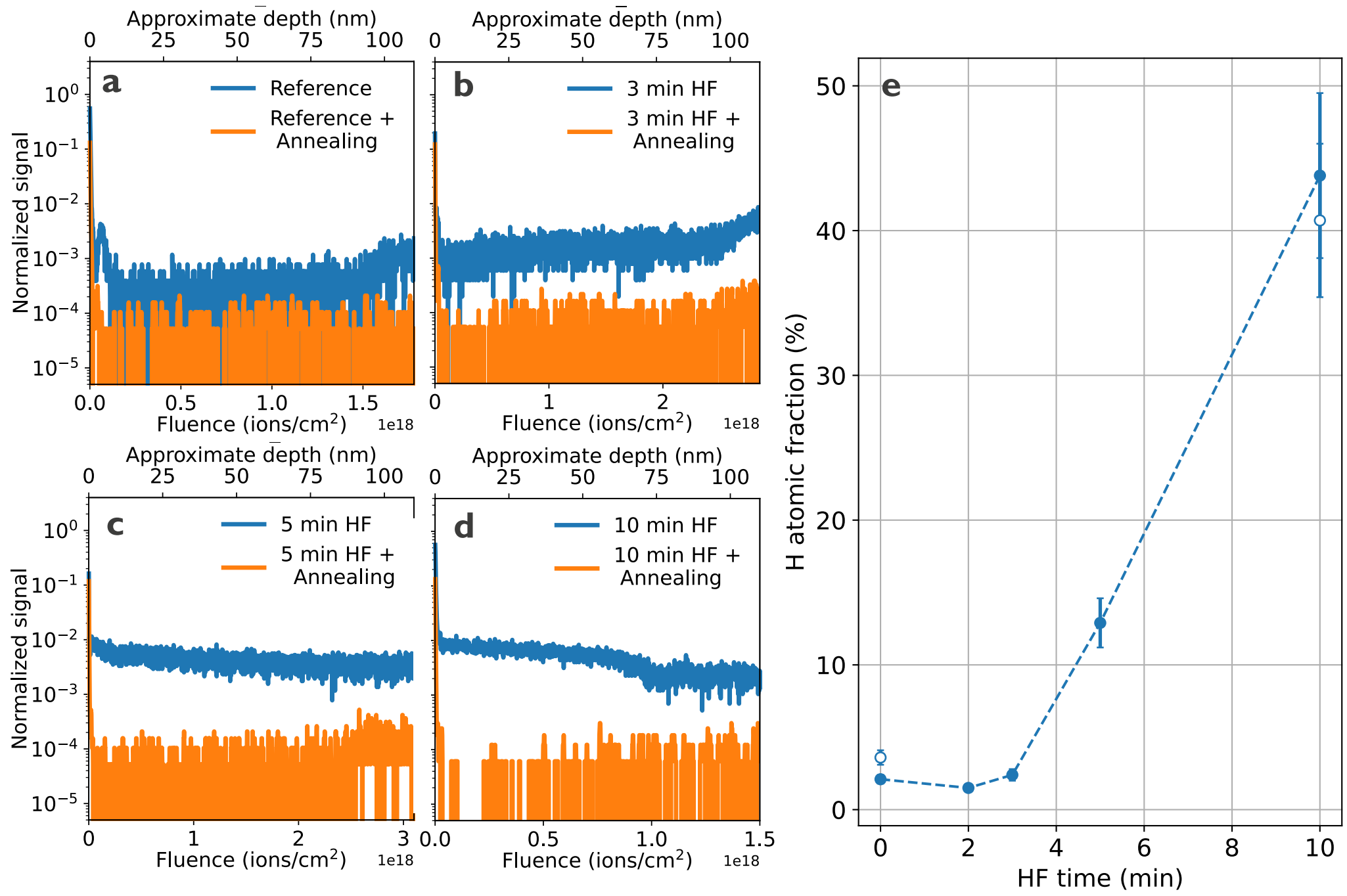

*Figure 2: ToF-SIMS hydrogen signal (normalized counts per second) as a function of fluence or approximate depth for **a** reference, **b** 3 min, **c** 5 min and **d** 10 min HF exposure. Results from samples subjected to a subsequent annealing are plotted in orange color. Silicon signal (not shown here) starts at an approximate depth of 100 nm. **e** Energy Recoil Detection Analysis (ERDA) results showing hydrogen atomic fraction as a function of HF exposure time. Empty symbols present additional measurements on a separate set of reference and 10 min HF samples to test reproducibility. Dashed line is a guide to the eye. Error bars represent atomic fraction uncertainties extracting from fitting ERDA spectra. Measurements are performed on the Ta surface hundreds of micrometers away from patterned edges.*

Superconducting transition temperature measurements show that Ta films exposed to HF for 10 minutes do not exhibit superconductivity, whereas films subjected to shorter exposures remain superconducting (see Figure S9 in Supporting Information). Furthermore, given that the only known superconducting tantalum hydride forms under extreme pressure (197 GPa),[31] we hypothesize that any tantalum hydride precipitates remain metallic at cryogenic temperatures and lead to ohmic loss. These findings suggest that excessive hydrogen incorporation can severely degrade superconducting properties, potentially leading to increased energy loss in superconducting qubits and reduced coherence times.

To eliminate the detrimental effects of hydrogen incorporation, an annealing procedure is employed to fully desorb hydrogen from the α-Ta film.[29] This approach is similar to previously demonstrated methods for hydrogen removal in bulk Nb films.[32] The annealing process of HF-treated samples is carried out at 500 °C for one hour under high vacuum (~$10^{-8}$ mbar). The removal of hydrogen from the film is confirmed by ToF-SIMS, where the $H^+$ signal in all annealed samples after HF treatment returns to background levels (orange datapoints in Figure 2**a,b,c,d**). The annealing parameters were not optimized, and it is likely that lower values would yield similar results. While prolonged HF exposure strongly affects the metal surface roughness, high-vacuum annealing does not further modify the surface (see Supporting Information). Notably, annealing restores superconductivity in the 10 min HF-treated sample, as indicated by the superconducting transition at 4.05 K (see Figure S10).

In addition to hydrogen, fluorine contamination could also contribute to the degradation of superconducting properties and affect the microwave performance of the resonator.

However, ToF-SIMS spectra of $F^-$ (see Figure S6) show that fluorine levels remain nearly identical across all samples, both after HF treatment and subsequent annealing. This consistency indicates that fluorine does not play a role in altering the superconducting properties of the film. Instead, the observed changes in our Ta samples can be attributed primarily to hydrogen, highlighting its dominant influence.

## Chemical analysis

A more comprehensive understanding of the chemical changes and surface oxidation in tantalum films resulting from HF treatment and subsequent high-vacuum annealing, is obtained through XPS analysis on the Ta top surface, hundreds of micrometers away from the patterned structures. Binding energy spectra show two Ta4f doublets (Figure 3**a**). The first at 28.0 eV and 26.2 eV, corresponds to Ta4$f_{5/2}$ and Ta4$f_{7/2}$ peaks of $Ta_2O_5$ and the second doublet at 22.6 eV and 20.7 eV corresponds to the metallic Ta4$f_{5/2}$ and Ta4$f_{7/2}$ peaks, respectively. The ratio of the $Ta_2O_5$/Ta peak amplitudes decreases with HF exposure duration, indicating that the HF treatment removes excess Ta oxide from the surface (see Supporting Information and Figure S8 for more details).

Interestingly, a significant shift of the metallic Ta doublet is observed with increasing HF duration. In the 10-minute HF sample, the Ta4$f_{5/2}$ and Ta4$f_{7/2}$ peaks shift from 22.6 eV to 23.3 eV and from 20.7 eV to 21.4 eV, respectively, compared to the reference sample. This is clearly visible in the binding energy shift, calculated as a binding energy difference between the $Ta_2O_5$ and the metallic Ta4$f_{7/2}$ peaks (Figure 3**c**). To confirm that this shift is not a measurement artifact, the measurements are repeated on two additional sets of samples, which consistently show the same trend (see Supporting Information). The observed binding energy shift is not related to charging effects, as this would result in the opposite effect, where instead of metallic, the oxide peak would gradually shifts to higher energies, and the metallic peak would remain unaffected.[33] The above observation, together with the ToF-SIMS data, suggests that the shift in the Ta metallic peaks arises from a hydrogen-rich chemical environment around Ta atoms, either due to hydrogen diluted in the film or the formation of tantalum hydrides.

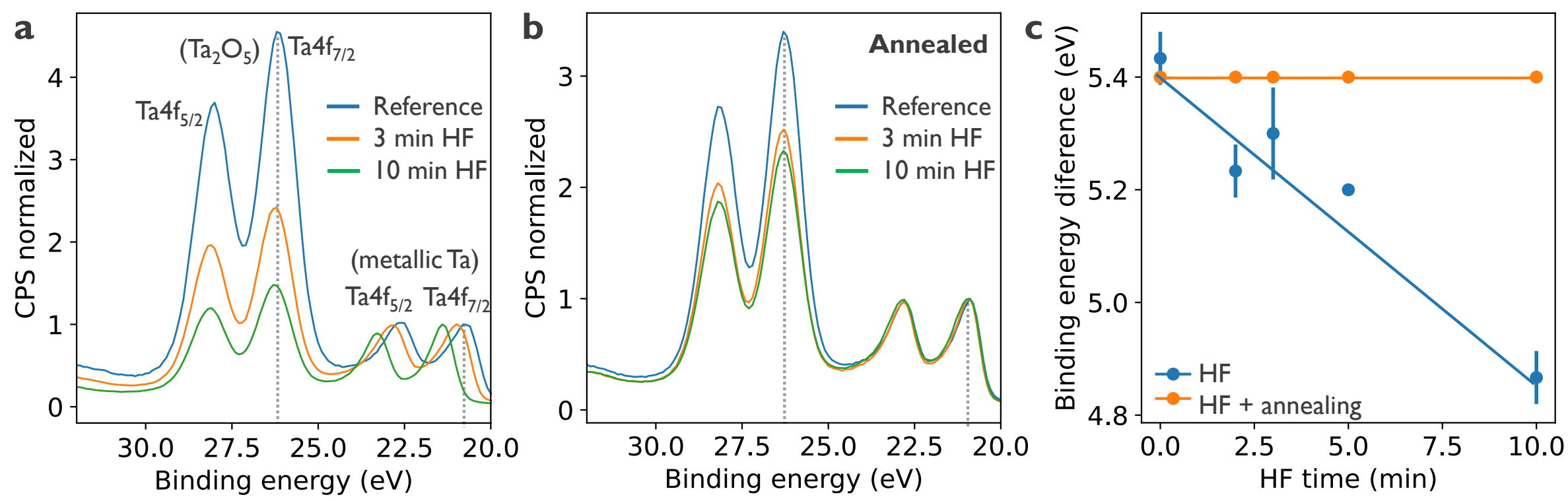

*Figure 3: **a** Ta binding energy spectrum (normalized counts per second) obtained with surface XPS for selected reference, 3 min HF and 10 min HF samples. **b** similar to **a** but, with an additional annealing step. Spectra are normalized to the intensity of the metallic Ta4$f_{7/2}$ peak and charge shift is calibrated using C1s peak (see Methods section in Supporting Information). **c** binding energy difference between the oxide and the metallic Ta4$f_{7/2}$ peaks indicated with dashed vertical lines for the HF treated samples with and without the high vacuum annealing. Error bars represent the standard deviation calculated from three independent measurements of the HF-only treated samples.*

Moreover, the fact that only the metallic peaks are affected, while the oxide peaks remain unchanged, strongly suggests that hydrogen is absorbed exclusively within the tantalum metal and not the oxide. This is further supported by XPS data from annealed samples, where the metallic Ta doublet (Figure **3b**) or its corresponding binding energy difference (orange points in Figure 3**c**) show no shift with increased HF exposure time. This indicates that hydrogen desorption fully restores the electronic environment of the tantalum lattice to that of the untreated reference sample. To our knowledge, this is the first reported observation of such a binding energy shift, since no reference for tantalum hydrides or hydrogen-rich tantalum films are present in well-known XPS databases.[34]

## Microwave loss

The impact of hydrogen incorporation in α-Ta film on microwave loss is evaluated by measuring the intrinsic quality factor ($Q_i$) of high-Q resonators in a dilution refrigerator at ~10 mK temperature (see Methods in Supporting Information for details). Resonators serve as effective proxies for superconducting qubits, as they are sensitive to many of the same microwave loss mechanisms while being easier and faster to fabricate and measure. Each sample contains eight coplanar-waveguide resonators with resonant frequencies evenly distributed between 4.2 and 7.8 GHz. All resonators share the same geometry, with a central trace width of $w$ = 24 μm and a gap of $s$ = 12 μm between the trace and the ground plane.

The resonator internal Q-factor characterization is performed on all resonators across differently processed samples to obtain statistically relevant results. For clarity, only single-photon level (low-power) $Q_{i,LP}$ and high-power $Q_{i,HP}$ datapoints in the linear regime (up to ~$10^8$ photons) are shown in Figure 4**a**. A pair of $Q_{i,LP}$ - $Q_{i,HP}$ datapoints is connected by a vertical line for each resonator. Resonators exposed to HF for 1 min and 2 min exhibit approximately twice the median $Q_{i,LP}$ (~4M) compared to the reference sample ($Q_{i,LP}$, ~ 2M), consistent with our previous findings.[9] Resonators exposed to HF for 3 min show a notable decrease in $Q_i$ factors down to $Q_{i,LP}$ ~ 1M, and a negligible power dependence. This correlates with an increased amount of absorbed hydrogen in the 3-min HF sample (~4% atomic fraction, see Figure 2**e**). In samples exposed to HF for 5 min and 10 min, resonances show drastically lower $Q_i$ values of ~$10^4$ and ~200, respectively. These resonances are particularly challenging to detect and analyze due to the large mismatch between internal and coupling quality factors. Notably, annealing restores internal Q-factors for all HF-treated samples, including 5- and 10-minute samples (orange points in Figure 4**a**), indicating that the observed microwave loss is correlated to the amount of hydrogen in α-Ta films.

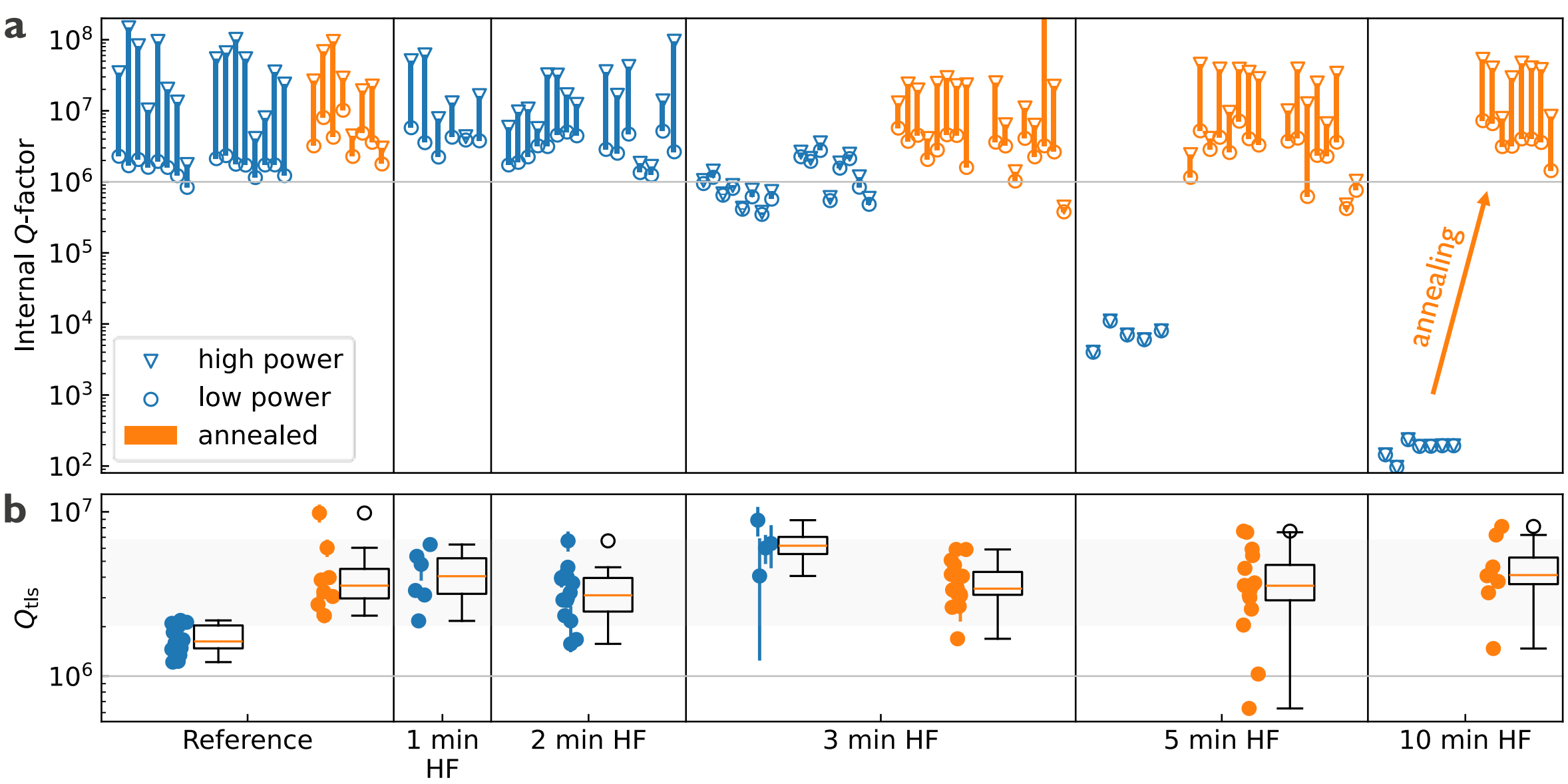

*Figure 4: High-Q resonator measurement results. **a** high power (triangle) and low power (circle) internal Q-factor datapoints are connected with a vertical line for each resonator. Datapoints corresponding to resonators from different samples are separated with a larger gap in horizontal direction than datapoints corresponding to the same sample. Resonators are horizontally ordered according to their frequency. Samples subject to UHV annealing are plotted with the orange color. While time dependence of low power Qi was measured for selected resonators (not shown), their standard deviation is generally lower than resonator-to-resonator variation. **b** Extracted $Q_{TLS}$ for all measured resonators are presented on a scatter plot grouped by different sample treatments. Error bars indicate the uncertainty in the $Q_{TLS}$ parameter obtained from the nonlinear fit. Datapoints with unreasonable uncertainty, i.e., where the uncertainty exceeded the parameter's value, were excluded from the analysis. Box plot is added to each scatter plot, where the box extends from the lower to upper quartile values of the data, with a line at the median. The whiskers extend from the box and show the range of the data up to 1.5x the inter quartile range (IQR). Flier datapoints beyond that range are considered outliers.*

To distinguish power-dependent TLS loss from power-independent loss, all extracted $Q_i$ in the linear regime (powers below nonlinear Duffing behavior) as a function of photon number are fitted using a TLS model (Equation S2) for different samples (Figure 4**b**). This analysis reveals that TLS loss ($1/Q_{TLS}$) is not affected by HF exposure up to 3 minutes, while the power-independent loss increases with longer HF exposure affecting both high-power and low-power $Q_i$ (Figure 4**a**). Furthermore, after annealing, the 3-, 5- and 10-minute samples exhibit restored power dependence of $Q_i$, with TLS loss comparable to that of unannealed 1- and 2-minute HF samples with $Q_{TLS}$ ~ 4M. For the summary of all fitting parameters see Figure S12 and subsection Statistical Analysis in Supporting Information for more details.

It is worth noting that the 5- and 10-minute HF-treated and annealed samples exhibit Ta surface roughness approximately ten times higher than the reference sample (Figure 1**b,c** and Figure S2). This increased roughness is expected to raise the participation ratio of the metal-air interface.[4] However, since the $Q_{TLS}$ of these samples remains comparable to those with shorter HF exposure and smoother surfaces, this suggests that tantalum oxide at the metal-air interface is not the dominant loss mechanism in $\alpha$-Ta resonators. Furthermore, the silicon-metal interface is unlikely to be affected by annealing, as the 500 °C annealing temperature is similar to the deposition temperature.[9]

It is also interesting to note that $Q_{TLS}$ of the annealed reference sample is approximately two times higher than that of the non-annealed reference samples and comparable to 1- and 2-min HF samples (Figure 4**b**). This could be a result of modification of the native silicon oxide layer present on the reference sample, reduction of tantalum sub-oxides (Figure S8) or removal of microwave TLS loss related to hydrogen[35] in the annealed reference sample (Figure 2**a**). However, further investigation is needed to determine the exact mechanism behind the TLS loss reduction during annealing.

Additional insight into the impact of hydrogen on superconducting circuits can be obtained by analyzing resonator performance at high power levels, beyond the linear regime. At microwave powers above ~$10^8$ photons, superconducting high-Q resonators typically exhibit Kerr-type nonlinear behaviour.[36–40] Interestingly, the 3-minute HF sample displays this nonlinear response at much lower photon numbers, around ~$10^3$. In addition, the onset of nonlinear Kerr-type frequency shift, is accompanied by an increase in microwave loss at higher photon numbers (empty blue circles in Figure 5**a**). The power dependence of microwave loss can be well modelled with a two-photon loss mechanism, which is typically associated with quasiparticle heating.[36,37] Further supporting this interpretation, the measured scattering parameter $S_{21}$ data in the nonlinear regime exhibit an elliptical shape in the IQ plane (Figure 5**c**), a consequence of strong power-dependent loss.[36] Fitting a nonlinear model to the measured scattering parameters allows to extract Kerr ($K_{nl}$) and two-photon ($\gamma_{nl}$) parameters, which are consistent across all resonators on this sample and range between 0.5 kHz and 3 kHz (Figure S11). For further details on the nonlinear model, see the Nonlinear resonator modeling section in the Supporting Information.

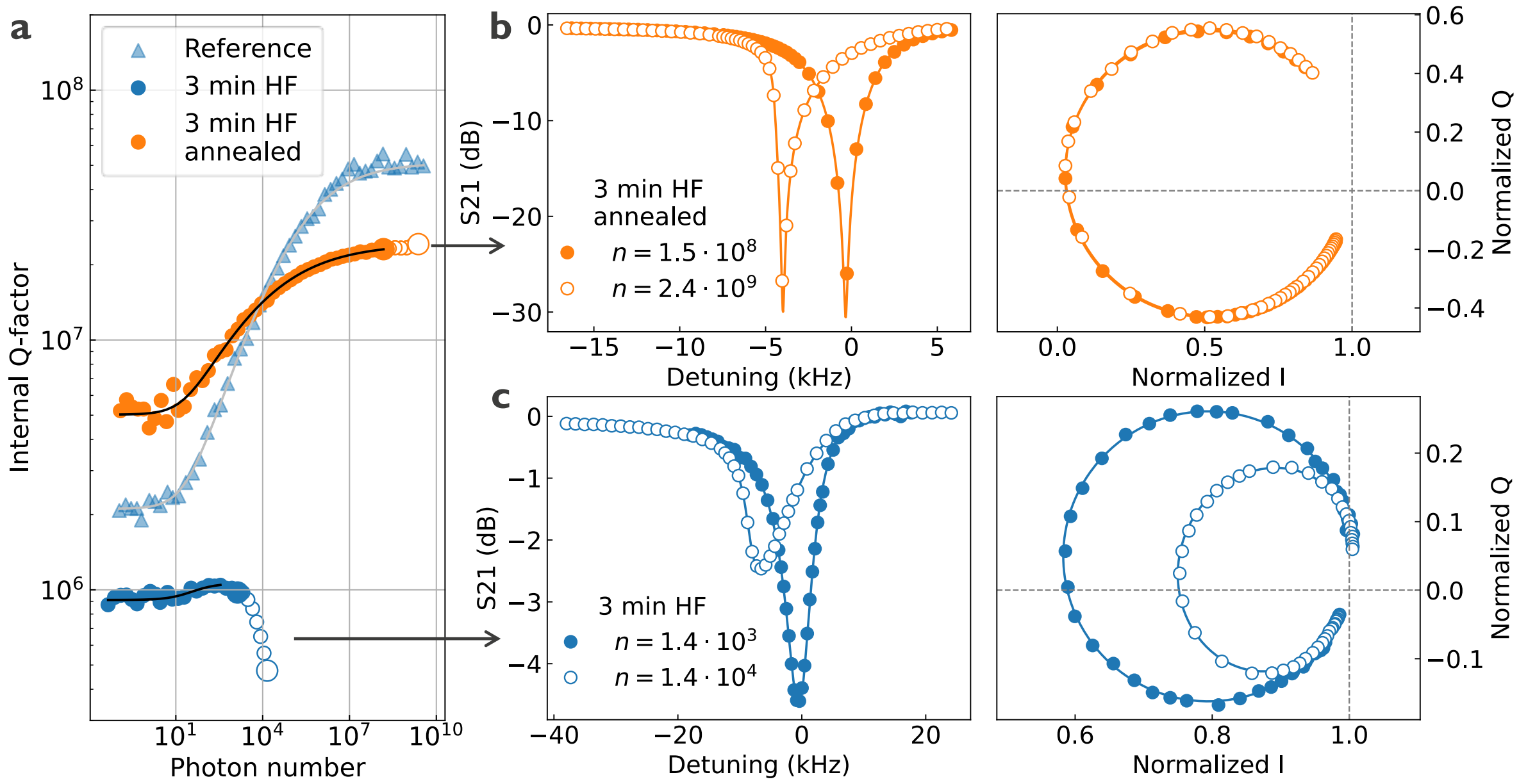

*Figure 5: Power dependent internal Q-factor measurement. **a** Power or photon number dependent internal Q-factor for resonator R1 on samples exposed to different treatments indicated in the legend. Filled markers are obtained by fitting the scattering parameter data to a linear model (Eq. S1) and empty markers are obtained by using a nonlinear model (Eq. S8). **b** Scattering parameters as a function of frequency for sample placed in HF for 3 min and received subsequent annealing treatment. The two traces correspond to a linear and nonlinear regime indicated with a photon number in the legend and with large markers in panel **a**. **c** The same as for panel **b**, but on 3 min HF sample that was not thermally treated. In the latter one, non-circular IQ trace is clearly visible, indicating the presence of nonlinear two-photon energy loss in high-Q resonators.*

After annealing the 3-min HF sample, the high-power behavior of the $Q_i$ recovers, exhibiting the expected Kerr-type nonlinearity at expected high photon numbers (~$10^8$ and above). The $S_{21}$ data in the IQ plane retains a fully circular trace, and the extracted Kerr parameter decreases significantly to $K_{nl}$ = 5 x $10^{-6}$ Hz at $n$ = 2.4x$10^9$ (Figure 5**b**). The same behavior is observed in the reference, the 1-minute and 2-minute HF samples, and in all annealed samples. This value is also consistent with previously reported high-power characteristics of superconducting high-Q resonators.[38,39]

The observed power-independent loss and nonlinear behavior in α-Ta samples exposed to HF for 3 minutes and longer might be a explained by ohmic losses due to the formation of non-superconducting tantalum hydride, or by off-resonant TLS relaxation-damping loss and TLS heating—mechanism previously associated with hydrogen in thin $Si_3N_4$ films.[35] While the exact origin of the loss remains to be determined, this work reveals a previously unreported hydrogen-related loss in α-Ta films, analogous to that observed in Nb resonators.[22]

## Conclusions

In summary, we demonstrate that prolonged HF exposure of α-Ta superconducting films introduces a measurable amount of hydrogen into the α-Ta bcc crystal structure, as confirmed by ToF-SIMS, ERDA, and XPS. Additionally, extended HF exposure leads to increased power-independent microwave loss and subsequent suppression of the resonance response in high-Q resonators, which could be explained by the formation of non-superconducting tantalum hydride or hydrogen-induced relaxation dampening TLS loss. Nonlinear resonator analysis further supports this claim, revealing an anomalous two-photon loss mechanism and increased Kerr nonlinearity in samples exposed to HF for 3 minutes or longer. However, annealing at 500 °C in UHV effectively removes hydrogen and restores high-Q performance, bringing $Q_i$ values back to those of lightly HF-treated samples. Our results establish a link between hydrogen incorporation and microwave loss in $\alpha$-Ta films, offering a clear strategy to reverse hydrogen-induced degradation in Ta- and Nb-based superconducting quantum devices. This key insight unlocks the potential for more effective etching and cleaning processes in device fabrication, paving the way for enhanced reliability in next-generation superconducting qubits.

## Acknowledgements

The authors gratefully thank Paola Favia, Olivier Richard, Chris Drijbooms, Ilse Hoflijk, Thierry Conard, Céline Noël, Valentina Spampinato, Alexis Franquet and Ryan Leong for metrology support. This work was supported in part by the imec Industrial Affiliation Program on Quantum Computing. We thank L. Swenson, G. Marcaud and H.-Y. Hsu for insightful comments on this work.

## Conflict of Interest

The authors declare no conflict of interest.

## Data availability

The data that supports the findings of this study are available from the corresponding authors upon reasonable request.

## References

1. Acharya, R. *et al.* Quantum error correction below the surface code threshold. *Nature* 1–3 (2024) doi:10.1038/s41586-024-08449-y.
2. Beverland, M. E. *et al.* Assessing requirements to scale to practical quantum advantage. Preprint at https://doi.org/10.48550/arXiv.2211.07629 (2022).
3. Leon, N. P. de *et al.* Materials challenges and opportunities for quantum computing hardware. *Science* **372**, (2021).
4. Mueller, C., Cole, J. H. & Lisenfeld, J. Towards understanding two-level-systems in amorphous solids: insights from quantum circuits. *Rep. Prog. Phys.* **82**, 124501 (2019).
5. Siddiqi, I. Engineering high-coherence superconducting qubits. *Nat. Rev. Mater.* 1–17 (2021) doi:10.1038/s41578-021-00370-4.

6. Van Damme, J. *et al.* Advanced CMOS manufacturing of superconducting qubits on 300 mm wafers. *Nature* **634**, 74–79 (2024).
7. Place, A. P. M. *et al.* New material platform for superconducting transmon qubits with coherence times exceeding 0.3 milliseconds. *Nat. Commun.* **12**, 1779 (2021).
8. Wang, C. *et al.* Towards practical quantum computers: transmon qubit with a lifetime approaching 0.5 milliseconds. *Npj Quantum Inf.* **8**, 1–6 (2022).
9. Lozano, D. P. *et al.* Low-loss α-tantalum coplanar waveguide resonators on silicon wafers: fabrication, characterization and surface modification. *Mater. Quantum Technol.* **4**, 025801 (2024).
10. Crowley, K. D. *et al.* Disentangling Losses in Tantalum Superconducting Circuits. *Phys. Rev. X* **13**, 041005 (2023).
11. McLellan, R. A. *et al.* Chemical Profiles of the Oxides on Tantalum in State of the Art Superconducting Circuits. *Adv. Sci.* **n/a**, 2300921 (2023).
12. Shi, L. *et al.* Tantalum microwave resonators with ultra-high intrinsic quality factors. *Appl. Phys. Lett.* **121**, 242601 (2022).
13. Marcaud, G. *et al.* Low-Loss Superconducting Resonators Fabricated from Tantalum Films Grown at Room Temperature. Preprint at https://doi.org/10.48550/arXiv.2501.09885 (2025).
14. Bal, M. *et al.* Systematic improvements in transmon qubit coherence enabled by niobium surface encapsulation. *Npj Quantum Inf.* **10**, 1–8 (2024).
15. Pritchard, P. G. & Rondinelli, J. M. Suppressed paramagnetism in amorphous Ta$_2$O$_{5-x}$ oxides and its link to superconducting qubit performance. Preprint at https://doi.org/10.48550/arXiv.2410.13160 (2024).
16. Vargas, P., Miranda, L. & Lagos, M. Diffusion Coefficient of Hydrogen in Niobium and Tantalum*. *Z. Für Phys. Chem.* **164**, 975–983 (1989).

17. Miranda, L., Vargas, P., Cerón, H. & Lagos, M. Hydrogen diffusion in tantalum. *Phys. Lett. A* **131**, 445–448 (1988).

18. Wipf, H. Solubility and diffusion of hydrogen in pure metals and alloys. *Phys. Scr. T* **94**, (2001).

19. Romanenko, A., Barkov, F., Cooley, L. D. & Grassellino, A. Proximity breakdown of hydrides in superconducting niobium cavities. *Supercond. Sci. Technol.* **26**, 035003 (2013).

20. Knobloch, J. The "Q disease" in Superconducting Niobium RF Cavities. *AIP Conf. Proc.* **671**, 133–150 (2003).

21. Sung, Z. *et al.* Direct observation of nanometer size hydride precipitations in superconducting niobium. *Sci. Rep.* **14**, 26916 (2024).

22. Torres-Castanedo, C. G. *et al.* Formation and Microwave Losses of Hydrides in Superconducting Niobium Thin Films Resulting from Fluoride Chemical Processing. *Adv. Funct. Mater.* **34**, 2401365 (2024).

23. Lee, J. *et al.* Discovery of Nb hydride precipitates in superconducting qubits. Preprint at https://doi.org/10.48550/arXiv.2108.10385 (2021).

24. Murthy, A. A. *et al.* Potential Nanoscale Sources of Decoherence in Niobium based Transmon Qubit Architectures. Preprint at http://arxiv.org/abs/2203.08710 (2022).

25. Asakawa, T., Nagano, D., Denda, S. & Miyairi, K. Evaluation of Hydrogen in Tantalum Thin Films Using Secondary Ion Mass Spectrometry. *Jpn. J. Appl. Phys.* **47**, 649 (2008).

26. Verjauw, J. *et al.* Investigation of Microwave Loss Induced by Oxide Regrowth in High-Q Niobium Resonators. *Phys. Rev. Appl.* **16**, 014018 (2021).

27. Altoé, M. V. P. *et al.* Localization and Mitigation of Loss in Niobium Superconducting Circuits. *PRX Quantum* **3**, 020312 (2022).

28. Christensen, C., De Reus, R. & Bouwstra, S. Tantalum oxide thin films as protective coatings for sensors. in *Technical Digest. IEEE International MEMS 99 Conference. Twelfth*

*IEEE International Conference on Micro Electro Mechanical Systems (Cat. No.99CH36291)* 267–272 (IEEE, Orlando, FL, USA, 1999). doi:10.1109/MEMSYS.1999.746832.

29. San-Martin, A. & Manchester, F. D. The H-Ta (hydrogen-tantalum) system. *J. Phase Equilibria* **12**, 332–343 (1991).

30. Schober, T. & Wenzl, H. The systems NbH(D), TaH(D), VH(D) : Structures, phase diagrams, morphologies, methods of preparation. in *Hydrogen in Metals II* (eds. Alefeld, G. & Völkl, J.) vol. 29 11–71 (Springer Berlin Heidelberg, Berlin, Heidelberg, 1978).

31. He, X. *et al.* Superconductivity Observed in Tantalum Polyhydride at High Pressure. *Chin. Phys. Lett.* **40**, 057404 (2023).

32. Barkov, F., Romanenko, A., Trenikhina, Y. & Grassellino, A. Precipitation of hydrides in high purity niobium after different treatments. *J. Appl. Phys.* **114**, 164904 (2013).

33. Lizarbe, A. J., Major, G. H., Fernandez, V., Fairley, N. & Linford, M. R. Insight note: X-ray photoelectron spectroscopy (XPS) peak fitting of the Al 2p peak from electrically isolated aluminum foil with an oxide layer. *Surf. Interface Anal.* **55**, 651–657 (2023).

34. Justin Gorham. NIST X-ray Photoelectron Spectroscopy Database - SRD 20. National Institute of Standards and Technology https://doi.org/10.18434/T4T88K (2012).

35. Mittal, S. *et al.* Annealing reduces ${\mathrm{Si}}_{3}{\mathrm{N}}_{4}$ microwave-frequency dielectric loss in superconducting resonators. *Phys. Rev. Appl.* **21**, 054044 (2024).

36. Thomas, C. N., Withington, S., Sun, Z., Skyrme, T. & Goldie, D. J. Nonlinear effects in superconducting thin film microwave resonators. *New J. Phys.* **22**, 073028 (2020).

37. Yurke, B. & Buks, E. Performance of Cavity-Parametric Amplifiers, Employing Kerr Nonlinearites, in the Presence of Two-Photon Loss. *J. Light. Technol.* **24**, 5054–5066 (2006).

38. Frasca, S. *et al.* NbN films with high kinetic inductance for high-quality compact superconducting resonators. *Phys. Rev. Appl.* **20**, 044021 (2023).

39. Swenson, L. J. *et al.* Operation of a titanium nitride superconducting microresonator detector in the nonlinear regime. *J. Appl. Phys.* **113**, 104501 (2013).

40. Anferov, A., Suleymanzade, A., Oriani, A., Simon, J. & Schuster, D. I. Millimeter-Wave Four-Wave Mixing via Kinetic Inductance for Quantum Devices. *Phys. Rev. Appl.* **13**, 024056 (2020).

# Supporting Information

## Reversing Hydrogen-Related Loss in a-Ta Thin Films for Quantum Device Fabrication

*Daniel P. Lozano,* Massimo Mongillo, Bart Raes, Yann Canvel, Shana Massar, Ananthapadmanabha M. Vadiraj, Tsvetan Ivanov, Rohith Acharya, Jacques Van Damme, Joris Van de Vondel, Danny Wan, Anton Potočnik,* and K. De Greve*

## Contents

# Methods

Multiple sample sets were prepared and characterized in this study to validate the effects of HF treatment on patterned Ta films. Resonator samples that had previously undergone HF treatment and annealing were analyzed using microwave characterization and STEM measurements. Microwave characterization was conducted within one day of the HF treatment, whereas STEM measurements were performed 2–3 months later, after the samples had been exposed to ambient conditions. A separate sample set was used for AFM, XPS, and ToF-SIMS measurements, performed both after HF treatment and after annealing. All measurements were conducted within two months of the HF treatment. Additionally, two more sample sets were analyzed with XPS to confirm that the $Ta4f_{5/2}$ and $Ta4f_{7/2}$ shift was caused by the HF treatment. Finally, a separate sample set was used for ERDA measurements. All HF treatments were performed at different times.

## Scanning Transmission Electron Microscopy (STEM)

STEM was performed on samples coated with spin-on carbon (SOC) layer. Lamellae with thickness of <50 nm were cut with focused ion beam (FIB) using Helios 450. For this study, HAADF-STEM was used to investigate metal-air, substrate-air and substrate-metal interfaces.

## Time-of-Flight Secondary Ion Mass Spectrometry (ToF-SIMS)

ToF-SIMS measurements were performed using a TOFSIMS NCS instrument from ION-TOF GmbH. Negative ion profiles were measured in a dual beam configuration using a $Bi^+$ (15 keV) gun for analysis and a $Cs^+$ (500 eV) gun for sputtering, while positive ion profiles were measured in a dual beam configuration using a $Bi^+$ (15 keV) gun for analysis and a $O_2^+$ (500 eV) gun for sputtering.

## Atomic Force Microscopy (AFM)

The AFM measurements were performed using the system ICON PT equipped with Nanoscope V in a tapping mode configuration. The tip used was OCML-AC160TS. Both Ta and Si surfaces were scanned across 2x2 $\mu m^2$ areas with a resolution of 2 nm/pixel.

## X-ray Photoelectron Spectroscopy (XPS)

The XPS measurements were carried out in Angle Integrated mode using a QUANTES instrument from Physical electronics. The measurements were performed using a monochromatized photon beam of 1486.6 eV. A 100 micron-wide spot was used. Charge neutralization was used during this experiment. Sensitivity factors specific to each instrument were used to convert peak areas to atomic concentrations. A charge shift calibration was performed on the C1s peak from adventitious carbon (284.8eV) for all spectra.

## Elastic Recoil Detection Analysis (ERDA)

ERDA experiments utilized a primary ion beam of $^{35}Cl^{4+}$ accelerated to 8 MeV by a 2 MV tandem accelerator. The forward recoiled and scattered ions are detected with a Time of Flight – Energy (ToF-E) telescope. The telescope has a length of 755.4 mm and is installed at a forward scattering angle of 40°. The sample tilt is at 15°. Scattered Cl was used for Ta. Recoil signals were used for the other elements. The reported concentrations refer to atomic fractions: at%.

## High-Q resonator measurements

Hanger-type $\lambda/4$ coplanar-waveguide resonators were used to study microwave loss in Ta resonators subject to different post-fabrication treatments as described in the main text. Resonators have resonant frequencies equidistantly spread between 4.2 and 7.8 GHz with coupling Q-factors ranging between 0.2 M and 1.8 M. Central trace width is $w$ = 24 μm and gap between the trace and the ground plane is $s$ = 12 μm.

Subdies with 8 resonators are measured in a dilution refrigerator at ~10 mK using Keysight P5004 vector network analyzer (VNA). Detailed information on the experimental setup can be found in our previous work.[1–3] Frequency dependent complex scattering parameter $S_{21}$ was measured near the resonance frequency for all resonators as a function of applied microwave power (Figure 5**b**). Transmission $S_{21}$ scattering parameters are analyzed with an generalized linear resonance model derived for a hanger-type resonator geometry and asymmetric line shapes:[4,5]

$$S_{21} = A\, e^{i(\omega t_{\rm d}+\phi)} \left(1 - \frac{\delta_{\rm c}}{\delta_{\rm c}+\delta_{\rm i}} \frac{1-i\alpha}{1+2i\tilde{\Delta}}\right), \qquad (1)$$

In this expression $A$ is the amplitude of a line shape, $t_{\rm d}$ is electric delay and $\phi$ is the phase. $\alpha$ is a small line shape asymmetry factor ($\alpha \ll 1$). $\delta_{\rm i} = \kappa_{\rm i}/\omega_{\rm r} = 1/Q_{\rm i}$ is intrinsic loss, $\delta_{\rm c} = \kappa_{\rm c}/\omega_{\rm r} = 1/Q_{\rm c}$ is coupling loss and $\tilde{\Delta} = \frac{\omega-\omega_{\rm r}}{\kappa_{\rm i}+\kappa_{\rm c}}$ is normalized frequency detuning. $Q_{\rm i}$ and $Q_{\rm c}$ are intrinsic and diameter corrected coupling quality factors, respectively, as defined in Ref [6].

Intrinsic quality factor shows a characteristic power dependence modelled by the two-level-system loss which is expressed as[3,7]

$$\delta_{\rm i}(\bar{n}) = \frac{1}{Q_{\rm i}(\bar{n})} = \frac{1}{Q_{\rm TLS}} \frac{\tanh\left(\frac{\hbar\omega_{\rm r}}{2k_B T}\right)}{\left(1+\frac{\bar{n}}{n_{\rm c}}\right)^{\alpha}} + \delta_0, \qquad (2)$$

where $Q_{\rm TLS} = 1/(F\,\tan\delta_{\rm TLS})$, is the two-level-system quality factor. $Q_{\rm TLS}$ is a function of effective energy participation ratio $F$ of interfaces and where TLS defects reside and $\tan\delta_{\rm TLS}$, which is the intrinsic loss tangent for the material containing the TLS. $\delta_0$ is the contribution from power independent non-TLS loss, $n_{\rm c}$ is the critical photon number related to the saturation electric field of TLS and $\alpha$ is a phenomenological parameter accounting for geometric effects[8] and the deviation from the standard TLS model.[9] Furthermore, $\hbar$ is the Planck constant, $\omega_{\rm r}$ is resonator frequency, $k_{\rm B}$ is Boltzmann constant and $T$ is the temperature, comparable to the base temperature of the dilution refrigerator with $T_{\rm base}{\sim}10$ mK.

$\bar{n}$ is average photon number in the resonator calculated as[10] $\bar{n} = \frac{2P_{\rm in}}{\hbar\omega_r^2}\frac{\delta_c}{(\delta_c+\delta_i)^2}$, where $P_{\rm in}$ is microwave signal power at the input of the resonator. An effective attenuation of approximately 74 dB between the VNA and the resonator inside the dilution refrigerator was separately estimated using ac-Stark shift and $\chi$-shift measurements with a superconducting transmon qubit.[11]

## Statistical analysis

We perform a statistical analysis on the extracted $Q_{TLS}$ values from several resonators from differently treated samples (see main text). To compare $Q_{TLS}$ values across samples, we combine scatter plots with box plots in Figure 4**b** (plotted on a log scale). In each box plot the box spans the interquartile range (IQR), from the 25% to 75% percentile, horizontal line denotes the median, whiskers extend to the furthest datapoints within 1.5 x IQR and points outside this range are marked as outliers.

In the "Microwave loss" section we claim that the following samples exhibit comparable $Q_{TLS}$: (i) annealed reference (ii) 1min HF, (iii) 2min HF, (iv) 3min HF, (v) annealed 3min HF, (vi) annealed 5min HF and (vii) annealed 10min HF. To support this, we applied a one-way ANOVA across these seven groups, which contain 8, 6, 14, 4, 16, 16 and 8 $Q_{TLS}$ measurements, respectively (see Figure 4**b**). We emphasize that the sample sizes are small, especially in group (iv) where datapoints were removed due to excessive uncertainty in the nonlinear fit, so this analysis is indicative rather than conclusive.

Given that the log-transformed $Q_{TLS}$ distribution appears more normally distributed (Figure 4**b**), we conduct ANOVA on both the raw and log-transformed values using *scipy.stats.f_oneway* python package. Both analyses returned $p$-values above the 0.05 significance threshold, indicating no statistically significant differences in $Q_{TLS}$ among these seven samples (see Table 1), corroborating our claim.

*Table 1: One-way ANOVA results for $Q_{TLS}$ and log-transformed $Q_{TLS}$ across seven treatment groups specified in the text above.*

| | ***p*-value** | ***F*-statistics** |
|---|---|---|
| $Q_{TLS}$ values | 0.11 | 1.82 |
| $\log(Q_{TLS})$ values | 0.24 | 1.38 |

# Ta and $TaO_x$ etch rates

The etch rate of tantalum oxide was conducted by timed immersion of Ta resonators samples in 10 vol% HF solution. Using atomic force microscopy (AFM), the Ta film height difference was measured between the bottom of the Si resonator gap and the top of the Ta film before and after dipping the samples in HF. Since HF does not attack Si in the range of concentration and treatment durations used in this study, the height difference corresponds to the amount of film removed by the HF. While individual datapoints can be measured with high accuracy, two separate sample sets were measured to gain a better estimate of measurement uncertainty. The first set was immersed for 2, 3, 5, and 10 minutes, while the second set underwent immersion for 8, 9, 10, and 11 minutes. Figure S1 shows that the film thickness remains unchanged up to approximately 9 minutes, with a noticeable reduction visible for 10-minute duration. The 9-minutes step is considered to completely remove the tantalum oxide (~3.7 nm) and marks the beginning of tantalum metal etching. Therefore, we estimate the etching rate of tantalum oxide in a 10 vol% HF solution to be 3.7 nm/9 min = 0.4 nm/min, and the etching rate of the tantalum film to be approximately 15 nm/min. The measured removal of 35 nm of Ta at 10 min is consistent with STEM results shown in Figure 1**a**.

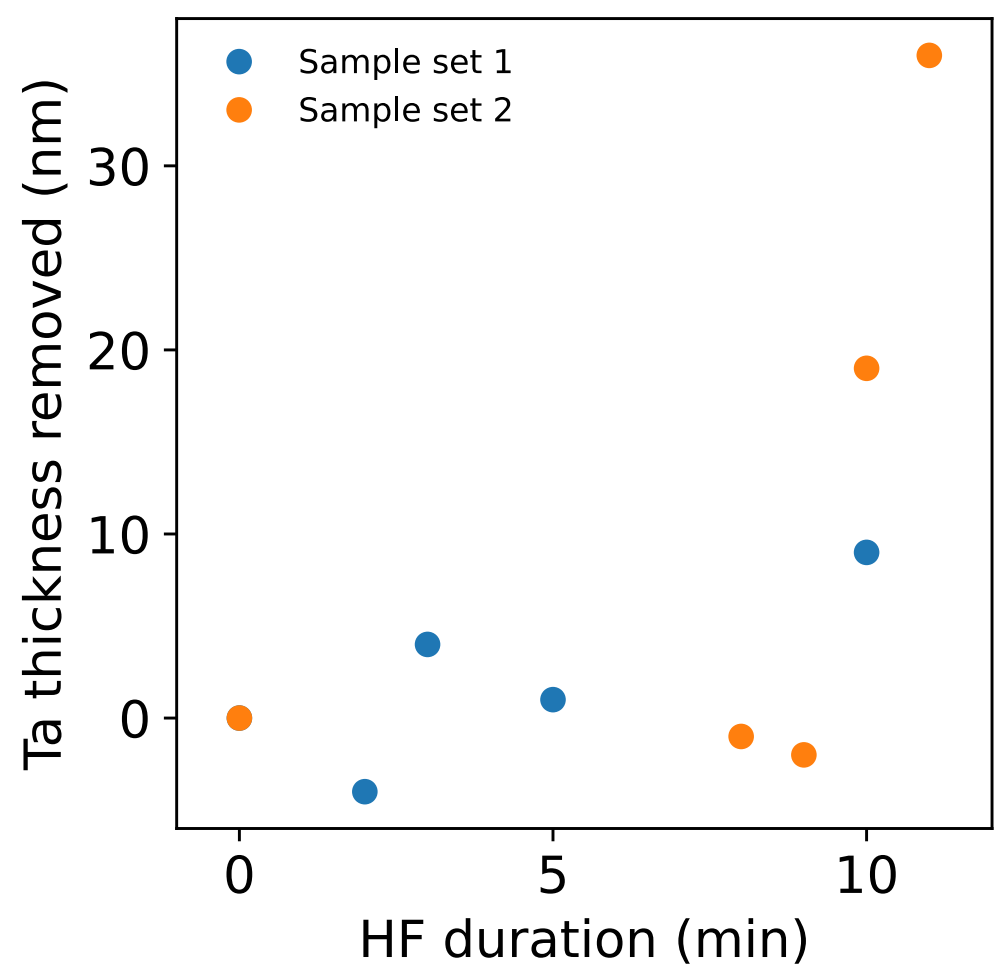

*Figure S1: Ta film height thickness removed as a function of HF treatment time.*

# Surface roughness (AFM)

AFM measurements were performed to analyze how the surface topology of Ta resonators evolves with HF treatment duration and annealing. Surface roughness was determined on both Si and Ta surfaces, with measurements taken after HF treatment and again after HF treatment followed by annealing.

The Ta surface height distribution and mean square roughness ($R_q$) remains consistent across the reference sample and those treated with HF for 2-3 minutes but shows significant increases in the 5- and 10-minute HF samples (Figure S2**a**). These surface characteristics persist after annealing at 500 °C (Figure S2**b**), indicating the thermal treatment does not alter the surface topology. In contrast, the Si surface exhibits no discernible pattern across any of the samples (Figure S2**c-d**). This confirms our expectations that neither HF treatment nor annealing affects the Si topology, and that any observed features can be attributed to the Si recess generated during the metal etch step in the samples. All the features discussed above can also be seen in the AFM maps shown in Figure S3**a-t**.

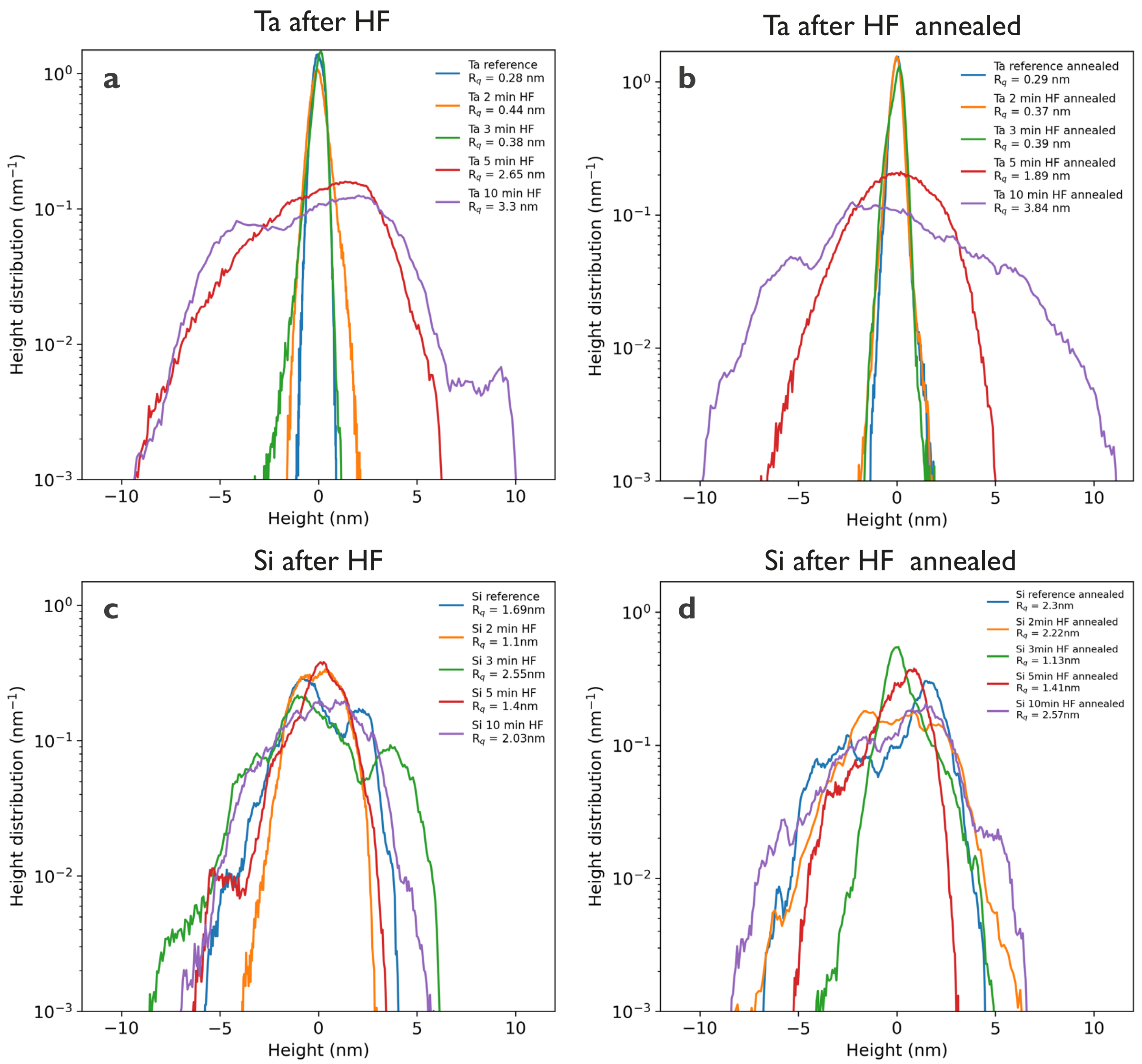

*Figure S2: Height distribution on the* ***a*** *the Ta surface after the HF treatment,* ***b*** *on the Ta surface after the HF treatment and annealing,* ***c*** *on the Si surface after the HF treatment and* ***d*** *on the Si surfacer after the HF treatment and the annealing for the reference (blue), 2 min HF (orange), 3 min HF (green), 5 min HF (red) and 10 HF samples d. The inset lists* $R_q$ *values for each of the samples.*

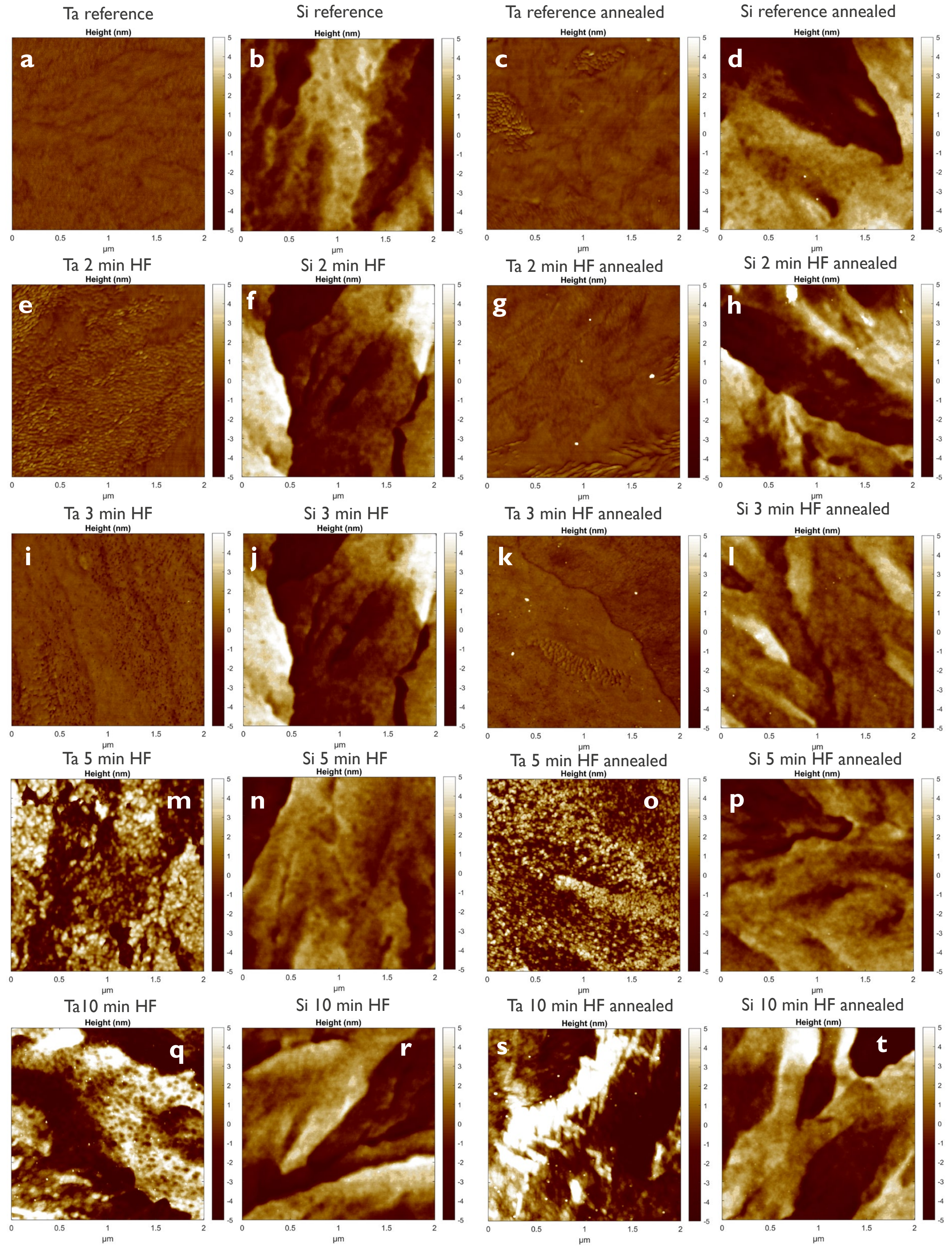

*Figure S3: **a-t** AFM scans for the Ta and Si surfaces after the HF treatment and after the HF treatment and annealing.*

# ToF-SIMS characterization

ToF-SIMS technique is used to detect hydrogen, fluorine and carbon and its distribution in the Ta films. Measurements were taken at two stages: after HF treatment and after subsequent annealing. Figure S4 displays the $H^+$ spectra alongside the $^{30}Si^+$ signal, which was added to clearly mark the interface between the Ta film and Si substrate. The Ta film thickness of 100 nm was identified at the point where the $^{30}Si^{+}$ signal begins to plateau.

We observe a distinct hydrogen peak at the Ta-Si interface in the 2 min HF and 3 min HF treated samples. In the 5 min and 10 min HF sample, the peak is not visible, likely due to the elevated hydrogen content throughout the film (see Figure S4). While variations in elemental signals can sometimes result from changes in ionization yield at material interfaces (e.g. from Ta to Si),[12] the weak hydrogen peak in the reference sample suggests a different origin. These observations support the claim that hydrogen initially enters the metal at the Ta-Si-air triple point. An alternative explanation is hydrogen incorporation at the Ta-Si interface from HF during silicon wafer cleaning prior to Ta deposition; however, such hydrogen will likely evaporate during high-temperature Ta deposition process at ~500°C and, if present, should contribute similarly across all samples.

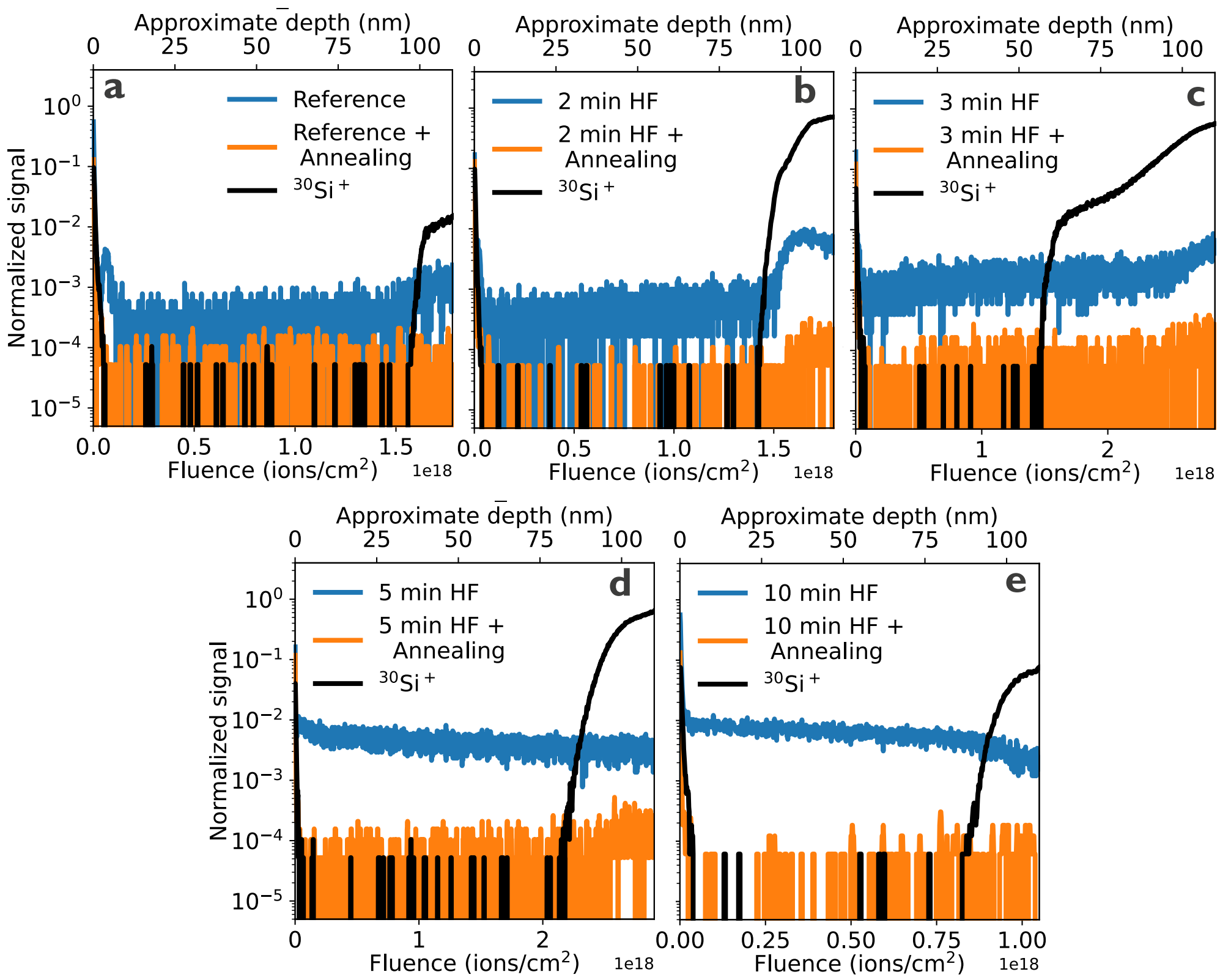

*Figure S4: **a-e** ToF-SIMS spectra for the $H^{+}$ after the HF treatment (blue) and after annealing (orange). The black line indicates the $^{30}Si^{+}$ signal. Increase of $H^{+}$ signal at the onset of $^{30}Si^{+}$ signal for samples with up to 3-minute exposure to HF could be related to changes in $H^{+}$ ionization yield or hydrogen gradient established due to its entrance through the triple point.*

We also monitor the normalized $C^{-}$ (Figure S5) and $F^{-}$ (Figure S6) signals to evaluate whether HF treatments influence the levels of these impurities. All samples exhibit comparable levels of both $C^{-}$ and $F^{-}$, regardless of HF treatment or annealing. This consistency is expected for C, as HF does not effectively remove it, and a thin layer of adventitious carbon inevitably forms upon air exposure, even if surface C is initially removed.

The uniform F signal across samples indicates that F does not influence the superconducting properties of the Ta films, and that the observed shifts in the metallic Ta $4f_{7}/_{2}$ and Ta $4f_{5}/_{2}$ XPS peaks are due to hydrogen incorporation rather than the presence of F.

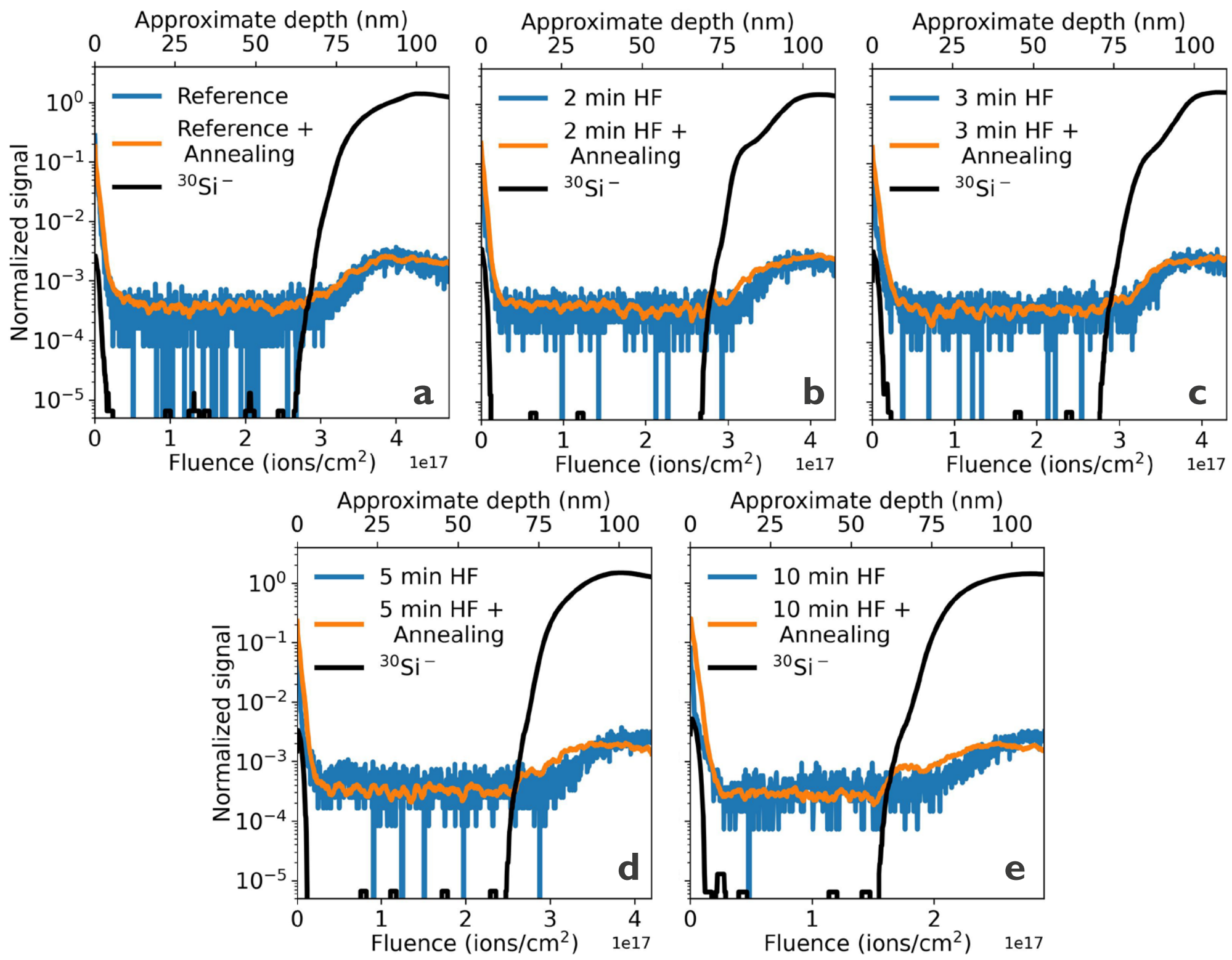

*Figure S5: **a-e** ToF-SIMS spectra for the $C^-$ after the HF treatment (blue) and after annealing (orange). The black line indicates the $^{30}Si^+$ signal.*

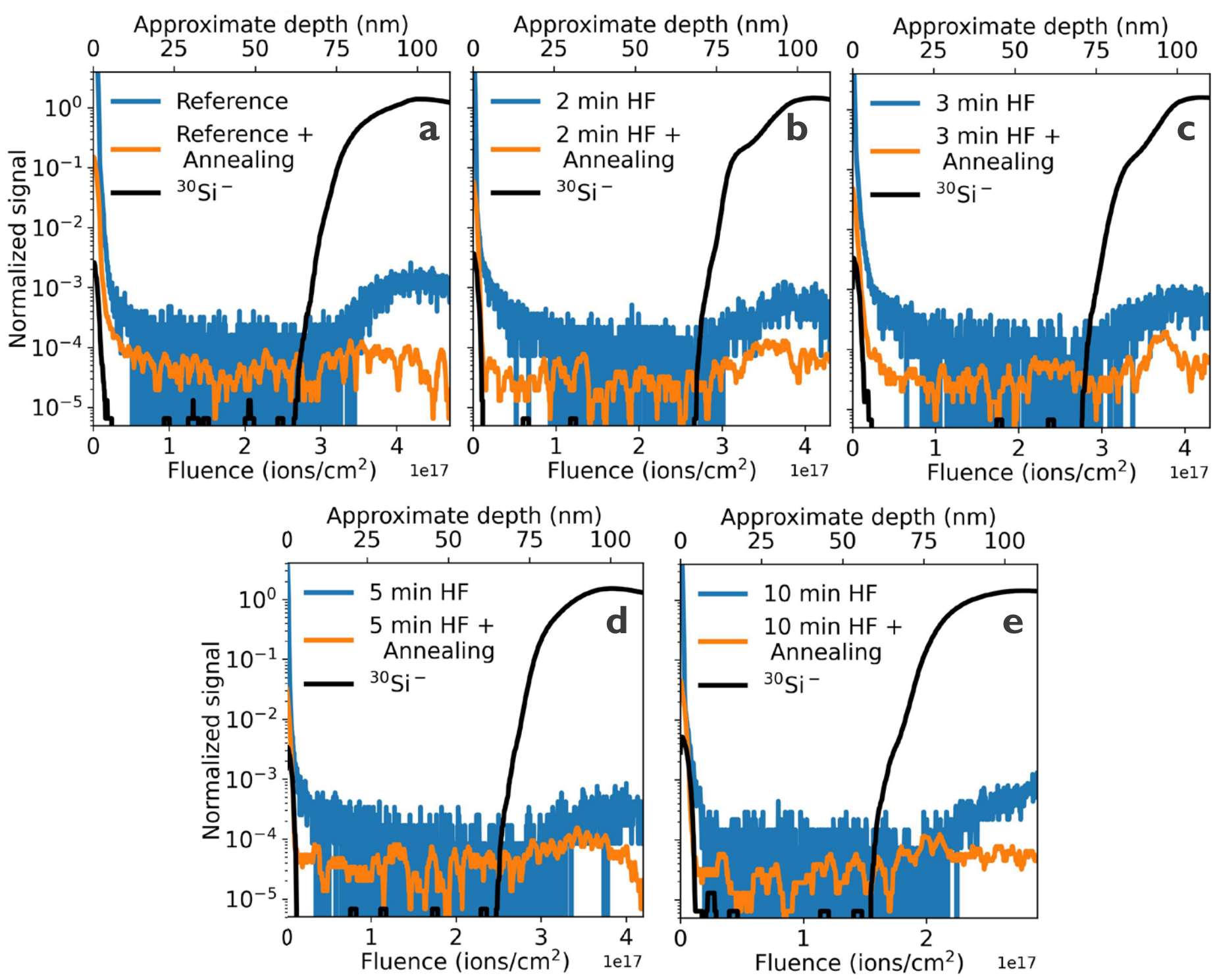

*Figure S6: **a-e** ToF-SIMS spectra for the $F^-$ after the HF treatment (blue) and after annealing (orange). The black line indicates the $^{30}Si^+$ signal.*

# XPS surface characterization

XPS measurements were performed to analyze the oxide composition on $\alpha$ -Ta film surface. Three distinct sample sets were examined in this study. In sample sets 1 and 3, measurements were performed only after HF treatment at varying durations, while sample set 2 (presented in Figure 3) underwent two measurement phases: first after HF treatment and second after subsequent annealing. To assess surface variability, three different points were measured on each sample after HF treatment, whereas for sample set 2, only a single point was analyzed post annealing.
The analysis reveals a progressive shift of the Ta4f metallic peaks toward higher binding energies with increasing HF duration, while the Ta4f peaks corresponding to $Ta_2O_5$ remained stable (Figure 3). Consequently, the binding energy separation between the $Ta4f_{7/2}$ peaks of $Ta_2O_5$ and metallic Ta decreases, as summarized in Figure S7. This shift is reversible, disappearing after 1-hour long UHV annealing at 500°C. Given its reproducibility across three sample sets and reversibility upon annealing, we attribute this behavior to hydrogen absorption and possibly tantalum hydride formation rather than charging effects. The presented data supports the findings in the main text (Figure 3) with additional measured samples, improving statistical reliability.

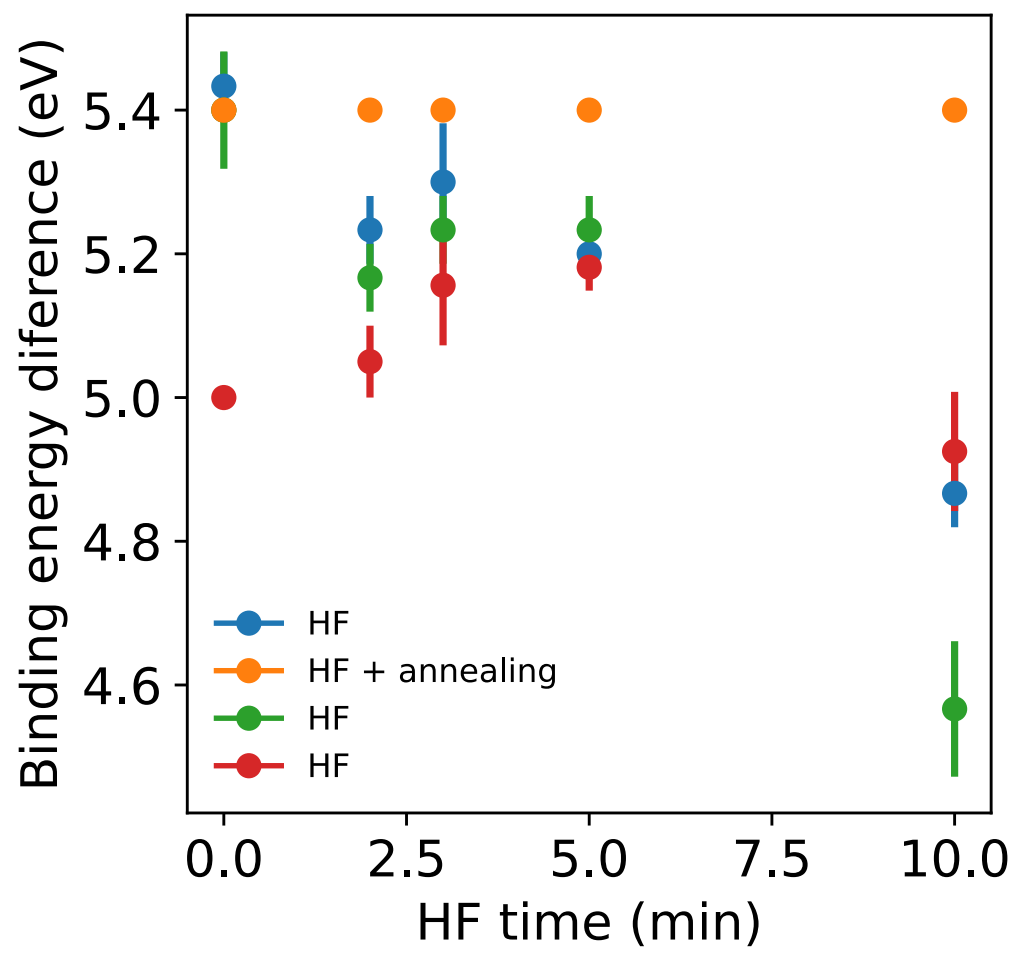

*Figure S7: Binding energy difference between the $Ta4f_{7/2}$ oxide and metallic peaks for sample set 1 after HF (blue), sample set 1 after HF and annealing (orange), sample set 2 after HF (green) and sample set 3 after HF (red). Markers and error bars represent the mean, and the standard deviation of the binding energy difference extracted from XPS spectra measured on 3 different points in the sample sets 1, 2 and 3 after the HF treatment (9 measurements in total). Sample set two correspond to the data shown in the main text.*

To quantitatively determine the amount of tantalum oxide $Ta_2O_5$, tantalum suboxide ($TaO_x$) and metallic tantalum components at the surface across sample sets 1, 2, and 3, the Ta4f spectra are fitted with mixed Gaussian-Lorentzian peaks[3,13] and the relative contributions of individual components are plotted in Figure S8 and summarized in Tables 2-5. While variations in total oxide content exist between sample sets, these differences remain within 15% for any given HF treatment duration. All sample sets exhibited the same trend, which is consistent with our previous results.[3] $Ta_2O_5$ is the predominant oxide present in all samples. The total amount of tantalum oxides ($Ta_2O_5$ + $TaO_x$) reduces from ~85% to ~70% within the first minute of HF exposure and remains largely unchanged with longer exposure times and subsequent annealing. This suggests that ~1-min HF treatment is sufficient to remove the excess tantalum oxide grown during oxygen-plasma-based resist strip,[3] while any further removal during prolonged HF exposure is offset by native oxide regrowth.

The atomic fractions of tantalum suboxides ($TaO_x$) are consistently higher for 2- and 3-minute exposure (blue bars in Figure S8), indicating that this pattern is a genuine phenomenon rather than an artifact of a single measurement set.

It has been hypothesized that surface roughness and pinholes allow the etching solution to access the buried suboxide layer and modify it.[14] While this is a possible mechanism, the reaction between $Ta_2O_5$ and HF is a complex four-step dissolution process[15] involving both sub-oxide and pentoxide intermediates that allow for the rearrangement and removal of oxygen atoms from the tantalum oxide structure. Further studies are needed to clarify how acid treatments can modify the suboxide layers in Ta films.

The amount of suboxides decreases significantly in all annealed samples, for instance, from 7% to 3% in the sample exposed to HF for 2 minutes (Figure S8). This suggests that suboxides undergo further oxidation, converting into $Ta_2O_5$. This transformation is expected, as $Ta_2O_5$ is the more thermodynamically stable oxide due to its higher valence state.[16] Metal suboxides like $TaO_x$ typically contain oxygen vacancies caused by deficiency of oxygen atoms in their crystal structure. Upon high-temperature annealing, $TaO_x$ undergoes complete oxidation, reducing oxygen vacancies as it converts into $Ta_2O_5$.[17]

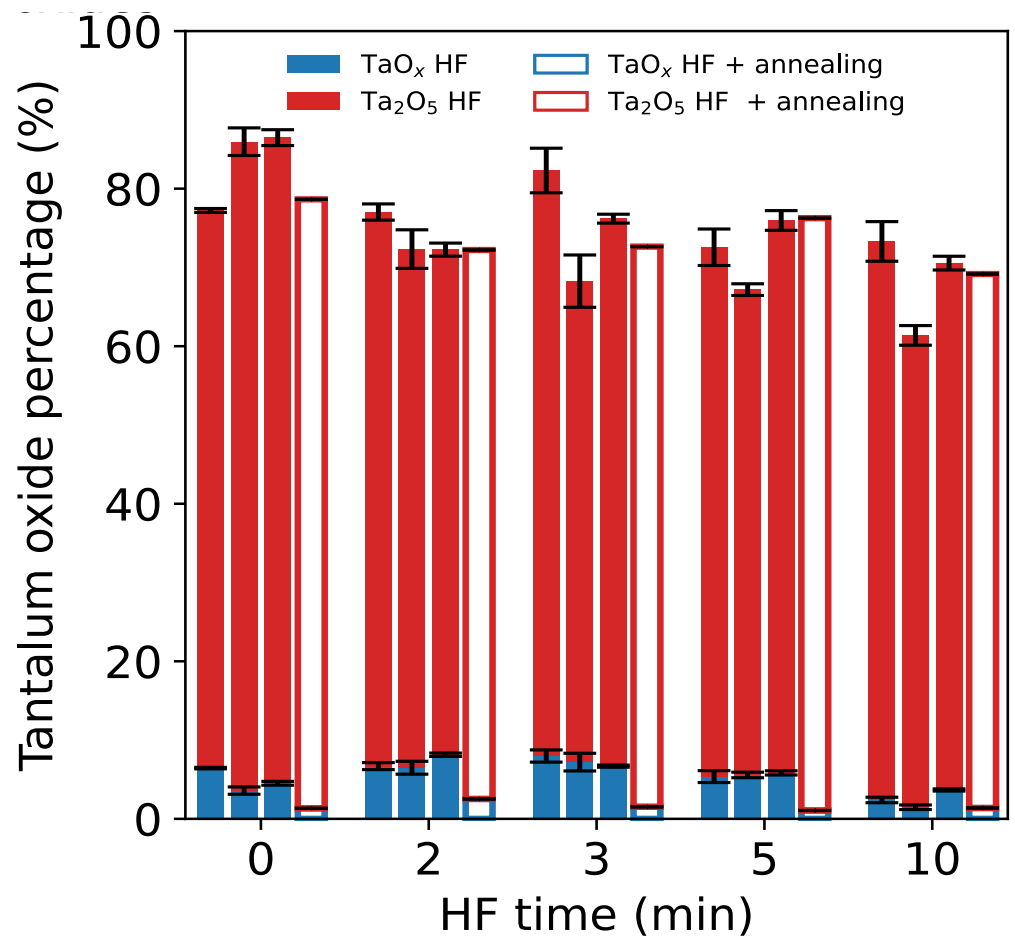

*Figure S8: Total tantalum oxide determined from fitting XPS spectra for different samples. Different samples are represented with different bars. The amount of suboxide $TaO_x$ is indicated by the bottom blue part and tantalum oxide $Ta_2O_5$ amount by the top red part of the bar plot. Annealed samples are indicated by empty bars. Error bars present standard deviation calculated from measurements on three different points for each bar (see Tables 2-5).*

*Table 2. Atomic concentrations extracted at different points on the sample surfaces after fitting the XPS spectrum for the Reference, 2 min HF treated, 3 min HF treated, 5 min HF treated, and 10 min HF treated sample set 1.*

| **Reference set 1** | **C (at%)** | **O (at%)** | **Si (at%)** | **Ta met (at%)** | **Ta sub.ox (at%)** | **Ta2O5 (at%)** |
|---|---|---|---|---|---|---|
| Point 1 | 28.05 | 51.50 | 2.53 | 2.37 | 0.83 | 14.72 |
| Point 2 | 27.61 | 51.16 | 1.66 | 2.68 | 0.86 | 16.02 |
| Point 3 | 24.45 | 53.13 | 2.28 | 2.75 | 0.91 | 16.49 |
| Mean | 26.65 | 51.92 | 2.12 | 2.59 | 0.87 | 15.73 |
| Standard deviation | 1.60 | 0.86 | 0.37 | 0.17 | 0.03 | 0.75 |
| **HF 2 min set 1** | | | | | | |
| Point 1 | 24.81 | 50.95 | 1.72 | 6.58 | 1.85 | 14.09 |
| Point 2 | 23.12 | 52.85 | 1.01 | 6.16 | 1.81 | 15.06 |
| Point 3 | 23.34 | 52.31 | 1.56 | 6.20 | 1.90 | 14.68 |
| Mean | 23.74 | 52.03 | 1.39 | 6.31 | 1.85 | 14.60 |
| Standard deviation | 0.75 | 0.80 | 0.30 | 0.19 | 0.04 | 0.40 |
| **HF 3 min set 1** | | | | | | |
| Point 1 | 20.97 | 55.75 | 1.41 | 5.27 | 1.49 | 15.12 |
| Point 2 | 21.00 | 54.79 | 1.27 | 5.44 | 1.53 | 15.97 |
| Point 3 | 21.18 | 55.13 | 0.92 | 5.38 | 1.50 | 15.89 |
| Mean | 21.05 | 55.22 | 1.18 | 5.36 | 1.51 | 15.66 |
| Standard deviation | 0.09 | 0.40 | 0.21 | 0.07 | 0.02 | 0.38 |
| **HF 5 min set 1** | | | | | | |
| Point 1 | 29.89 | 49.32 | 0.31 | 4.92 | 1.21 | 14.34 |
| Point 2 | 22.91 | 53.95 | 0.73 | 5.39 | 1.29 | 15.74 |
| Mean | 26.17 | 51.58 | 0.48 | 5.15 | 1.25 | 15.02 |
| Standard deviation | 3.49 | 2.32 | 0.21 | 0.24 | 0.04 | 0.70 |
| **HF 10 min set 1** | | | | | | |
| Point 1 | 23.60 | 52.10 | 1.85 | 7.31 | 0.87 | 14.27 |
| Point 2 | 22.85 | 52.08 | 1.70 | 7.87 | 0.81 | 14.69 |
| Mean | 23.22 | 52.09 | 1.77 | 7.58 | 0.84 | 14.48 |
| Standard deviation | 0.38 | 0.01 | 0.08 | 0.28 | 0.03 | 0.21 |

*Table 3. Atomic concentrations extracted at different points on the sample surface after fitting the XPS spectrum for the Reference + annealing, 2 min HF treated + annealing, 3 min HF treated + annealing, 5 min HF treated + annealing and 10 min HF treated + annealing sample set 1.*

| Reference + annealing set 1 | C (at%) | O (at%) | Si (at%) | Cs (at%) | Ta met (at%) | Ta sub.ox (at%) | Ta2O5 (at%) |
|---|---|---|---|---|---|---|---|
| Point 1 | 31.98 | 46.22 | 5.97 | 0.84 | 3.2 | 0.2 | 11.58 |
| **HF 2 min + annealing set 1** | | | | | | | |
| Point 1 | 29.23 | 47.63 | 3.96 | 0.55 | 5.18 | 0.47 | 12.99 |
| **HF 3 min + annealing set 1** | | | | | | | |
| Point 1 | 1.19 | 46.09 | 5.08 | 0.57 | 4.67 | 0.26 | 12.14 |
| **HF 5 min + annealing set 1** | | | | | | | |
| Point 1 | 43.35 | 38.57 | 3.58 | 0.35 | 3.36 | 0.15 | 10.64 |
| **HF 10 min + annealing set 1** | | | | | | | |
| Point 1 | 31.62 | 45.76 | 5.47 | 0.46 | 5.15 | 0.23 | 11.32 |

*Table 4. Atomic concentrations extracted at different points on the sample surfaces after fitting the XPS spectrum for the Reference, 2 min HF treated, 3 min HF treated, 5 min HF treated, and 10 min HF treated sample set 2.*

| Reference set 2 | C (at%) | O (at%) | Si (at%) | Ta met (at%) | Ta sub.ox (at%) | Ta2O5 (at%) |
|---|---|---|---|---|---|---|
| Point 1 | 24.37 | 53.78 | 0.91 | 3.00 | 0.76 | 17.17 |
| Point 2 | 37.45 | 43.27 | 1.74 | 2.42 | 0.61 | 14.51 |
| Point 3 | 24.42 | 53.33 | 0.58 | 3.03 | 0.79 | 17.85 |
| Mean | 28.14 | 49.88 | 0.97 | 2.80 | 0.72 | 16.44 |
| Standard deviation | 6.15 | 4.85 | 0.49 | 0.28 | 0.08 | 1.44 |
| **HF 2 min set 2** | | | | | | |
| Point 1 | 24.37 | 50.15 | 1.20 | 7.31 | 1.72 | 15.23 |
| Point 2 | 21.24 | 54.25 | 1.25 | 5.47 | 1.29 | 16.50 |
| Point 3 | 22.37 | 51.62 | 1.53 | 7.09 | 1.70 | 15.69 |
| Mean | 22.62 | 51.98 | 1.32 | 6.57 | 1.56 | 15.80 |
| Standard deviation | 1.29 | 1.70 | 0.15 | 0.82 | 0.20 | 0.52 |
| **HF 3 min set 2** | | | | | | |
| Point 1 | 22.28 | 51.61 | 1.20 | 8.42 | 1.91 | 14.58 |
| Point 2 | 22.46 | 53.73 | 1.08 | 5.99 | 1.40 | 15.34 |
| Point 3 | 21.96 | 51.46 | 0.76 | 8.90 | 2.04 | 14.88 |
| Mean | 22.23 | 52.26 | 0.99 | 7.66 | 1.76 | 14.93 |
| Standard deviation | 0.21 | 1.04 | 0.19 | 1.27 | 0.28 | 0.31 |
| **HF 5 min set 2** | | | | | | |
| Point 1 | 23.92 | 51.76 | 0.89 | 7.69 | 1.33 | 14.42 |
| Point 2 | 24.79 | 50.59 | 1.27 | 7.88 | 1.38 | 14.08 |
| Point 3 | 24.26 | 51.90 | 0.95 | 7.29 | 1.18 | 14.43 |
| Mean | 24.32 | 51.41 | 1.02 | 7.62 | 1.29 | 14.31 |
| Standard deviation | 0.36 | 0.59 | 0.17 | 0.25 | 0.08 | 0.16 |
| **HF 10 min set 2** | | | | | | |
| Point 1 | 24.73 | 50.58 | 1.46 | 8.99 | 0.34 | 13.90 |
| Point 2 | 23.88 | 50.48 | 1.03 | 9.98 | 0.48 | 14.15 |
| Point 3 | 24.02 | 52.00 | 0.61 | 8.51 | 0.26 | 14.60 |
| Mean | 24.30 | 50.53 | 1.23 | 9.47 | 0.40 | 14.02 |
| Standard deviation | 0.43 | 0.05 | 0.21 | 0.50 | 0.07 | 0.13 |

*Table 5. Atomic concentrations extracted at different points on the sample surfaces after fitting the XPS spectrum for the Reference, 2 min HF treated, 3 min HF treated, 5 min HF treated, and 10 min HF treated sample set 3.*

| **Reference set 3** | **C (at%)** | **O (at%)** | **Si (at%)** | **Ta met (at%)** | **Ta sub.ox (at%)** | **Ta2O5 (at%)** |
|---|---|---|---|---|---|---|
| Point 1 | 24.37 | 53.78 | 0.91 | 3.00 | 0.76 | 17.17 |
| Point 2 | 37.45 | 43.27 | 1.74 | 2.42 | 0.61 | 14.51 |
| Point 3 | 24.42 | 53.33 | 0.58 | 3.03 | 0.79 | 17.85 |
| Mean | 28.14 | 49.88 | 0.97 | 2.80 | 0.72 | 16.44 |
| Standard deviation | 6.15 | 4.85 | 0.49 | 0.28 | 0.08 | 1.44 |
| **HF 2 min set 3** | | | | | | |
| Point 1 | 24.37 | 50.15 | 1.20 | 7.31 | 1.72 | 15.23 |
| Point 2 | 21.24 | 54.25 | 1.25 | 5.47 | 1.29 | 16.50 |
| Point 3 | 22.37 | 51.62 | 1.53 | 7.09 | 1.70 | 15.69 |
| Mean | 22.62 | 51.98 | 1.32 | 6.57 | 1.56 | 15.80 |
| Standard deviation | 1.29 | 1.70 | 0.15 | 0.82 | 0.20 | 0.52 |
| **HF 3 min set 3** | | | | | | |
| Point 1 | 22.28 | 51.61 | 1.20 | 8.42 | 1.91 | 14.58 |
| Point 2 | 22.46 | 53.73 | 1.08 | 5.99 | 1.40 | 15.34 |
| Point 3 | 21.96 | 51.46 | 0.76 | 8.90 | 2.04 | 14.88 |
| Mean | 22.23 | 52.26 | 0.99 | 7.66 | 1.76 | 14.93 |
| Standard deviation | 0.21 | 1.04 | 0.19 | 1.27 | 0.28 | 0.31 |
| **HF 5 min set 3** | | | | | | |
| Point 1 | 23.92 | 51.76 | 0.89 | 7.69 | 1.33 | 14.42 |
| Point 2 | 24.79 | 50.59 | 1.27 | 7.88 | 1.38 | 14.08 |
| Point 3 | 24.26 | 51.90 | 0.95 | 7.29 | 1.18 | 14.43 |
| Mean | 24.32 | 51.41 | 1.02 | 7.62 | 1.29 | 14.31 |
| Standard deviation | 0.36 | 0.59 | 0.17 | 0.25 | 0.08 | 0.16 |
| **HF 10 min set 3** | | | | | | |
| Point 1 | 24.73 | 50.58 | 1.46 | 8.99 | 0.34 | 13.90 |
| Point 2 | 23.88 | 50.48 | 1.03 | 9.98 | 0.48 | 14.15 |
| Point 3 | 24.02 | 52.00 | 0.61 | 8.51 | 0.26 | 14.60 |
| Mean | 24.30 | 50.53 | 1.23 | 9.47 | 0.40 | 14.02 |
| Standard deviation | 0.43 | 0.05 | 0.21 | 0.50 | 0.07 | 0.13 |

# Superconducting $T_c$ measurements

Superconducting transition temperature is measured for samples exposed to HF for different durations using a four-probe measurement in an adiabatic demagnetization refrigeration (ADR) cryostat. The reference sample shows the lowest $T_c$ = 4.20 K, while samples exposed to HF between 2- and 5-minute exhibit somewhat higher values of $T_c \approx 4.25$ K (Figure S9). This increase could be a result of the reduced amount of total Ta oxide on the surface. Surprisingly, the sample that was exposed to HF for 10 minutes shows no superconducting transition down to 0.2 K. This might be due to the considerable amount of hydrogen absorbed or even $TaH_x$ formation (close to $\delta$ phase) in the metal which could render the entire film non-superconducting.

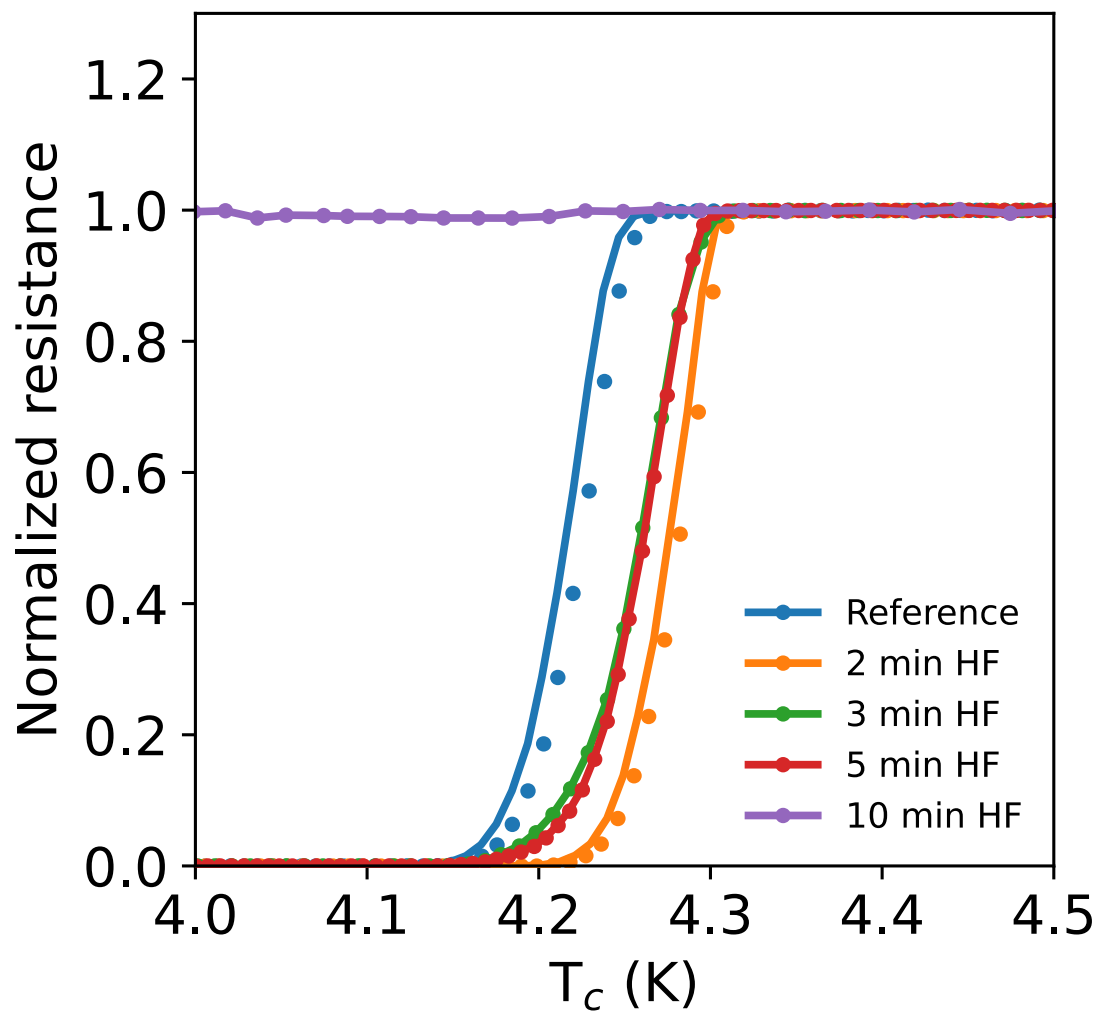

*Figure S9: Normalized resistance measurements as a function of temperature for Ta samples show superconducting transition at approximately $T_c$ = 4.25 K for samples exposed to HF of up to 5min, and no transition for 10-minute exposed sample down to 0.2 K (not shown).*

After annealing the 10-minute HF-treated sample, superconductivity is recovered with $T_c$ of approximately 4.05 K (Figure S10). Reduced $T_c$ compared to the reference value (4.20 K) is likely due to the sample being thinner (60 nm) than the reference sample (100 nm) as a result of Ta etching caused by HF treatment.[18]

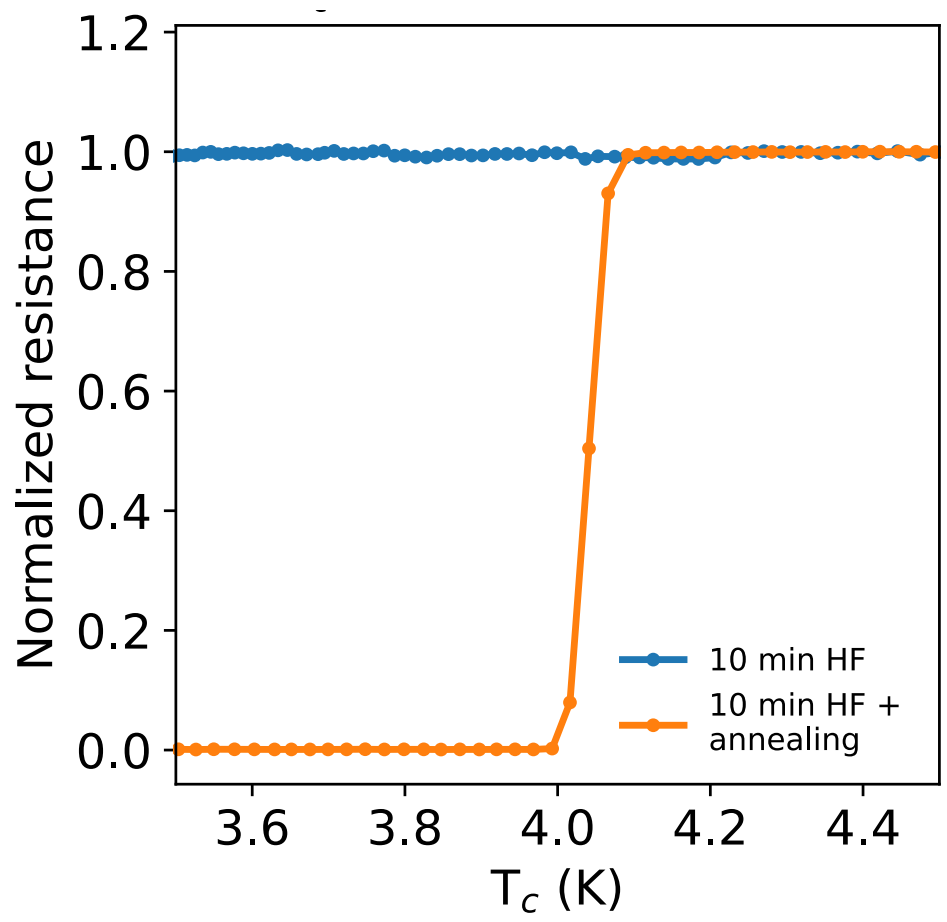

*Figure S10: Normalized resistance measurements as a function of temperature for Ta samples after 10 min HF (no superconducting transition down to 0.2 K, not shown) and after 10 min HF and annealing showing a superconducting transition at approximately $T_c$ = 4.05 K.*

# Nonlinear high-Q resonator modeling

In this section we describe the nonlinear model and a fitting strategy for measured S21 scattering parameter as a function of frequency for Ta resonators exposed to diluted HF for 3min (Figure 5). The nonlinear fitting model encompasses both nonlinear kinetic inductance ($L$) contribution as well as nonlinear ohmic loss ($R$) modelled with two-photon absorption contributions[19]

$$L = \Delta L\left(1 + \frac{I^2}{I_c^2}\right), \qquad R = \Delta R\left(1 + \frac{I^2}{I_c^2}\right), \tag{3}$$

where $I$ and $I_c$ are current and critical current, respectively. Both contributions arise when microwave current density through a superconducting transmission or a strip line becomes overcritical at its peak near the edges, in which case reduced Cooper pair density and increased quasiparticle density at the edges lead to current dependent kinetic inductance and resistance per unit length.[19,20]

The nonlinear Kerr effect is modelled with the following Hamiltonian,

$$\mathcal{H} = \hbar\omega_r a^\dagger a + \frac{\hbar}{2} K_{nl} a^\dagger a^\dagger a a, \tag{4}$$

and the two-photon loss is accounted for with the following equation of motion,

$$\dot{a} = -i\omega_r a - iK_{nl} a^\dagger a a - \frac{\kappa_i + \kappa_c}{2} a - \frac{\gamma_{nl}}{2} a^\dagger a a - \sqrt{\frac{\kappa_c}{2}} a_{in}. \tag{5}$$

In the above two expressions, $\hbar$ is the Planck constant divided by $2\pi$, $\omega_r$ is the resonator's fundamental frequency, $K_{nl}$ is the nonlinear Kerr parameter, $\gamma_{nl}$ is the nonlinear two-photon loss parameter, $\kappa_i$ is internal loss rate, $\kappa_c$ is coupling rate to the feedline, $a^\dagger$ and $a$ are creation and annihilation operators obeying bosonic commutation relation $[a, a^\dagger] = 1$ and $a_{in}$ is incoming drive photon flux.

Assuming coherent drive ($a(t) = ae^{-i\omega t}$) and multiplying both sides of Eq. (5) with its complex conjugate we arrive at the nonlinear equation for inter resonator photon number,

$$\frac{\kappa_c}{2}|a_{\text{in}}|^2 = n^3\left(K_{\text{nl}}^2 + \frac{\gamma_{\text{nl}}^2}{4}\right) + 2n^2\left[\frac{(\kappa_{\text{i}}+\kappa_{\text{c}})\gamma_{\text{nl}}}{4} - \Delta K_{\text{nl}}\right] + n\left[\frac{(\kappa_{\text{i}}+\kappa_{\text{c}})^2}{4} + \Delta^2\right], \tag{6}$$

where $n = a^\dagger a$ is intra-resonator photon number and $\Delta = \omega - \omega_r$ is frequency difference between the drive and the resonator's fundamental resonant mode.

Eq. (6) can be further simplified by normalizing the equation to[21,22]

$$\frac{1}{2} = \tilde{n}^3\left(\xi^2 + \frac{\eta^2}{4}\right) + 2\tilde{n}^2\left[\frac{\eta}{4} - \xi\tilde{\Delta}\right] + \tilde{n}\left[\frac{1}{4} + \tilde{\Delta}^2\right], \tag{7}$$

where

$$\tilde{n} = \frac{n}{|\tilde{a}_{\text{in}}|^2}, \quad \xi = \frac{|\tilde{a}_{\text{in}}|^2 K_{\text{nl}}}{\kappa_{\text{i}}+\kappa_{\text{c}}}, \quad \eta = \frac{|\tilde{a}_{\text{in}}|^2 \gamma_{\text{nl}}}{\kappa_{\text{i}}+\kappa_{\text{c}}}, \quad \tilde{\Delta} = \frac{\Delta}{\kappa_{\text{i}}+\kappa_{\text{c}}}, \quad |\tilde{a}_{\text{in}}|^2 = \frac{\kappa_c}{(\kappa_{\text{i}}+\kappa_{\text{c}})^2}|a_{\text{in}}|^2 .$$

When calculating the scattering parameter of a nonlinear resonator, the third order nonlinear Eq. (7) is solved first to obtain $\tilde{n}$ as a function of frequency $\tilde{\Delta}$, which is then used to expand the Lorentzian line shape[1] as follows:

$$S_{21} = A\, e^{i(\omega t_{\text{d}} + \phi)}\left(1 - \frac{\delta_{\text{c}}}{\delta_{\text{c}} + \delta_{\text{i}}}\frac{1 - i\alpha}{1 + \eta\tilde{n} + 2i(\tilde{\Delta} - \xi\tilde{n})}\right), \tag{8}$$

In the above expression $\xi\tilde{n} = \frac{K_{\text{nl}}\, n}{\kappa_{\text{i}}+\kappa_{\text{c}}}$ is the normalized nonlinear frequency shift and $\eta\tilde{n} = \frac{\gamma_{\text{nl}}\, n}{\kappa_{\text{i}}+\kappa_{\text{c}}}$ is the normalized nonlinear two-photon loss rate, with $K_{\text{nl}} \cdot n$ and $\gamma_{\text{nl}} \cdot n$ being nonlinear Kerr shift and two-photon loss rate, respectively, and $\alpha$ is a small line shape asymmetry factor ($\alpha \ll 1$).

The nonlinear frequency shift can lead to an asymmetric tilted Duffing response of $S_{21}(\omega)$ or a multi-solution response beyond the critical bifurcating point for $\xi < \xi_c$. Notably, $S_{21}(\omega)$ measurement points form a circle in an IQ plane across all parameter regimes, even for drive powers beyond the bifurcation point. Conversely, the nonlinear two-photon loss term results in a non-circular resonance curve in the IQ plane. This resonance curve takes on an approximately elliptical shape with the $S_{21}(\omega_{\text{r}})$ and asymptotic $S_{21}(\omega \rightarrow \pm\infty)$ points located at the co-vertices, the endpoints of the minor axis.[19,20] In the limit where $\xi \rightarrow 0$ and $\eta \rightarrow 0$, Eq. (8) converges to the linear generalized asymmetric Lorentzian line shape [Eq. (1)].[6]

The high-power scattering parameter data shown in Figure 5**c**,**d** is modelled by the following nonlinear fitting procedure implemented using the *lmfit* python package. Starting with initial set of fitting parameters, the photon number is computed for each frequency point using Eq. (7). These photon numbers and the remaining fitting parameters are used to compute the scattering parameter S21 as a function of frequency using Eq. (8). The nonlinear fitting is performed using the least square method (*leastsq*) from *scipy.optimize* package used by *lmfit*.

The nonlinear Kerr parameter ($K_{\text{nl}}$) and the nonlinear two-photon loss parameter ($\gamma_{\text{nl}}$) are obtained by repeating the described fitting routine for S21 spectra collected at higher microwave drive powers (empty markers in Figure S11**a,b**) and plotting nonlinear Kerr shift ($K_{nl} \cdot n$) and two-photon loss rate ($\gamma_{nl} \cdot n$) against the maximum photon number in a resonator. The nonlinear Kerr parameter and the two-photon loss parameter are finally extracted from the slope of the linear fit (Figure S11**c,d**), which are for Ta resonators exposed to diluted HF for 3 min summarized in Figure S11**e.** Surprisingly, $K_{nl}$ and $\gamma_{nl}$ have comparable values ranging from 0.5 to 3 kHz. This is possibly due to a common quasiparticle heating source.

[1] This line shape expression is adapted from Daniel Flanigan resonator package (https://github.com/danielflanigan) which is rewritten for convenient fitting with the lmfit python package.

Total internal quality factor combining linear intrinsic and the nonlinear loss $Q_i^{-1} = \frac{\kappa_i}{\omega_r} + \frac{\gamma_{nl} n}{\omega_r}$ extracted for each S21 spectrum at a given power agree well with a combination of a TLS intrinsic loss model (2) and a two-photon loss from a single parameter for each resonator presented in Figure S11**e** as shown with a dashed line in Figure S11**a,b**.

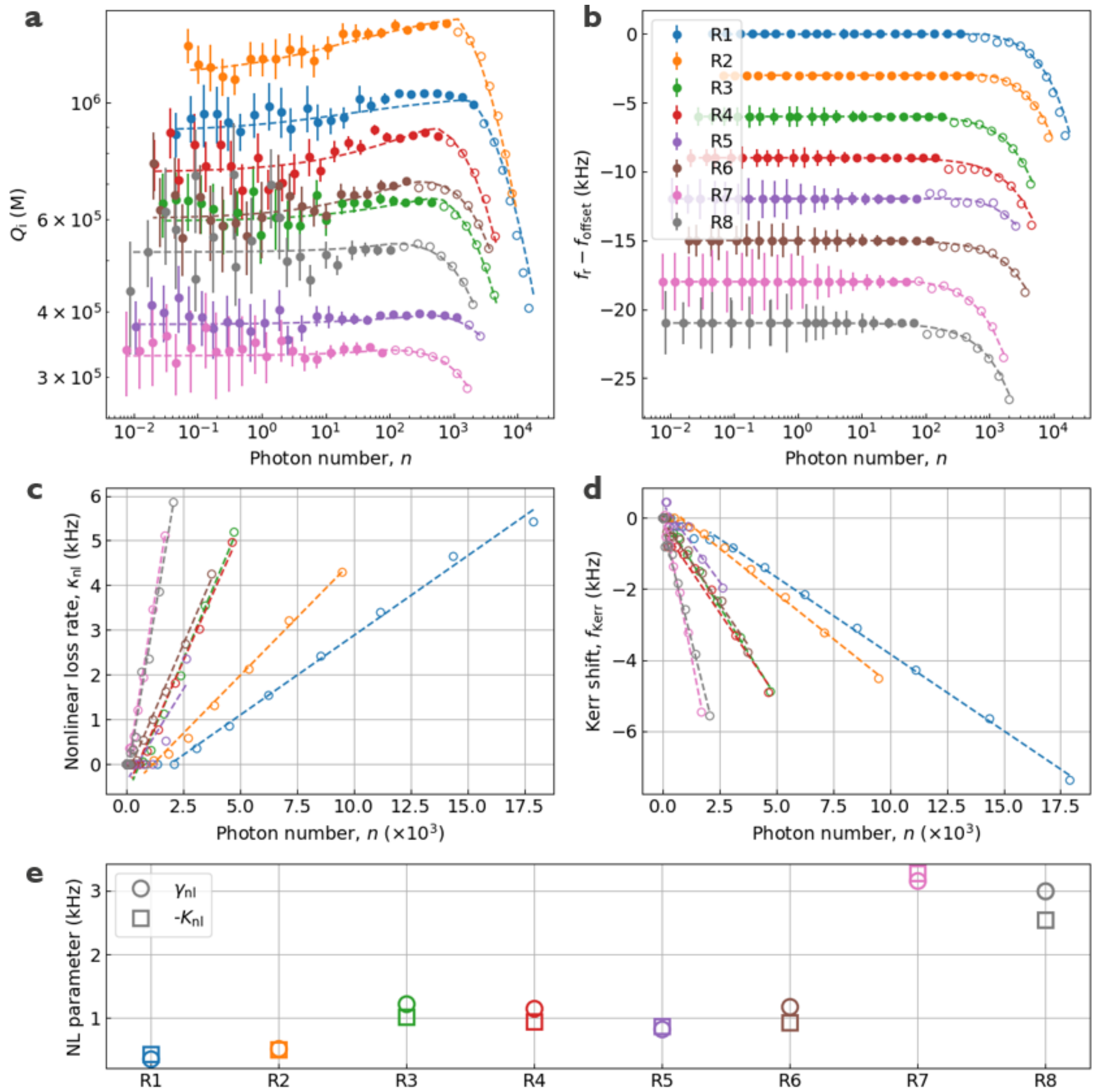

*Figure S11: Nonlinear analysis of Q-internal and resonator frequency shift for all 8 resonators on a Ta resonator sample exposed to 3min HF.* ***a*** *Internal Q-factor as a function of photon number in a resonator. Dashed lines correspond to total linear and nonlinear loss.* ***b*** *Resonator's frequency shift relative to their value at the lowest photon number as a function of the photon number in a resonator. For better visibility, data for each resonator is offset by 3 kHz. Dashed lines indicate nonlinear Kerr shift. Error bars in panels a and b represent the standard errors obtained from the nonlinear fit using the lmfit python package.* ***c*** *Nonlinear loss rate as a function of photon number and* ***d*** *Kerr shift as a function of the photon number.* ***e*** *nonlinear loss ($\gamma_{nl}$) and negative nonlinear Kerr ($-K_{nl}$) parameters extracted from slopes in subpanels* ***c*** *and* ***d*** *for each resonator on a chip.*

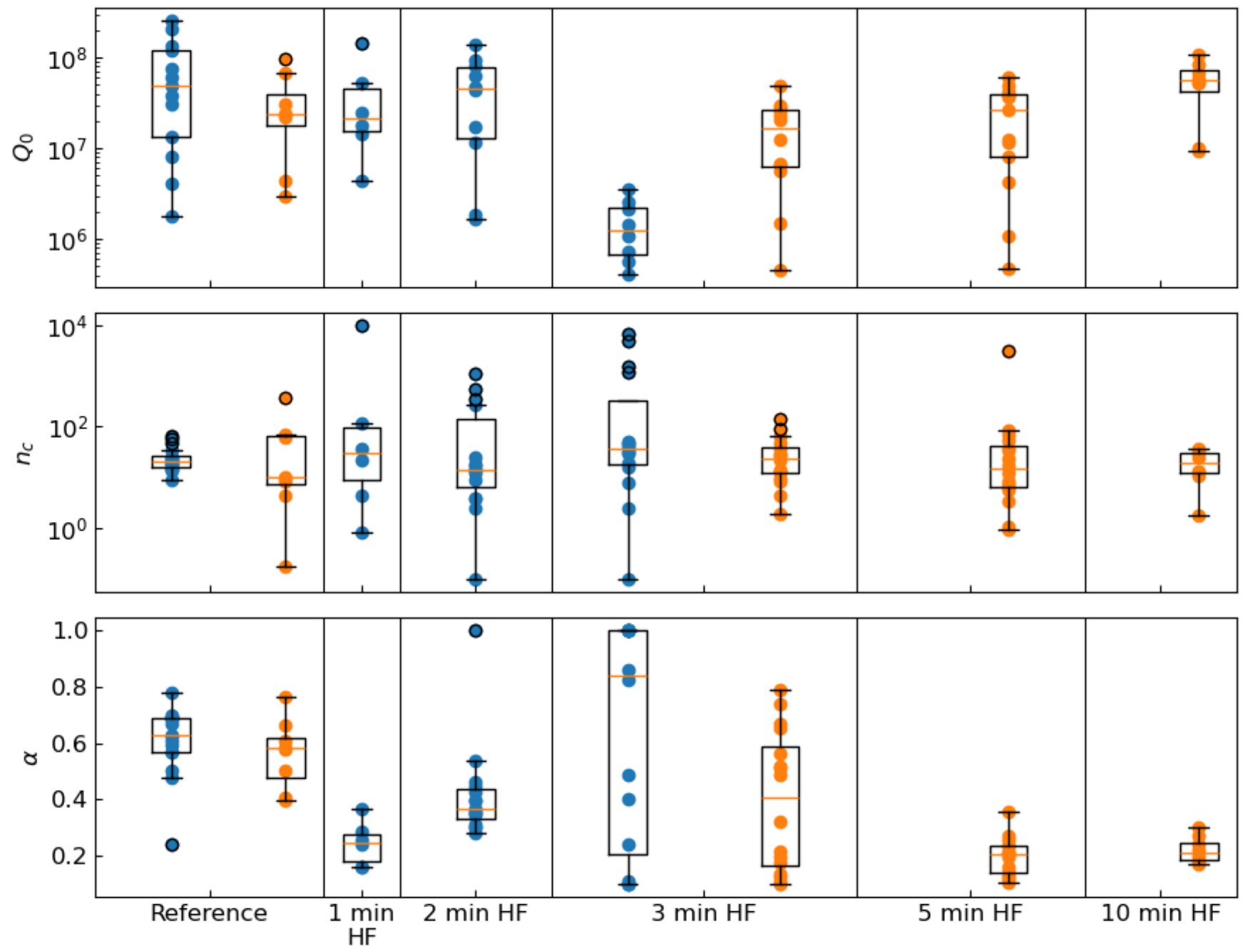

*Figure S12: Summary of TLS model fitting parameters for different resonators with different HF treatment. High-power $S_{21}$ measurements that exhibit nonlinear behavior were excluded from TLS model fitting Eq. (2) presented here. Parameters plotted here are: $Q_0$ power-independent loss, $n_c$ critical photon number, and $\alpha$ is the phenomenological parameter accounting for geometric effects.*

# References

1. Verjauw, J. *et al.* Investigation of Microwave Loss Induced by Oxide Regrowth in High-Q Niobium Resonators. *Phys. Rev. Appl.* **16**, 014018 (2021).
2. Van Damme, J. *et al.* Argon milling induced decoherence mechanisms in superconducting quantum circuits. Preprint at https://doi.org/10.48550/arXiv.2302.03518 (2023).
3. Lozano, D. P. *et al.* Low-loss α-tantalum coplanar waveguide resonators on silicon wafers: fabrication, characterization and surface modification. *Mater. Quantum Technol.* **4**, 025801 (2024).
4. Rieger, D. *et al.* Fano Interference in Microwave Resonator Measurements. *Phys. Rev. Appl.* **20**, 014059 (2023).

5. Khalil, M. S., Stoutimore, M. J. A., Wellstood, F. C. & Osborn, K. D. An analysis method for asymmetric resonator transmission applied to superconducting devices. *J. Appl. Phys.* **111**, 054510 (2012).
6. Probst, S., Song, F. B., Bushev, P. A., Ustinov, A. V. & Weides, M. Efficient and robust analysis of complex scattering data under noise in microwave resonators. *Rev. Sci. Instrum.* **86**, 024706 (2015).
7. Burnett, J., Bengtsson, A., Niepce, D. & Bylander, J. Noise and loss of superconducting aluminium resonators at single photon energies. *J. Phys. Conf. Ser.* **969**, 012131 (2018).
8. Wang, H. *et al.* Improving the coherence time of superconducting coplanar resonators. *Appl. Phys. Lett.* **95**, 233508 (2009).
9. Phillips, W. A. Tunneling states in amorphous solids. *J. Low Temp. Phys.* **7**, 351–360 (1972).
10. Bruno, A. *et al.* Reducing intrinsic loss in superconducting resonators by surface treatment and deep etching of silicon substrates. *Appl. Phys. Lett.* **106**, 182601 (2015).
11. Koch, J. *et al.* Charge-insensitive qubit design derived from the Cooper pair box. *Phys. Rev. A* **76**, 042319 (2007).
12. Priebe, A., Xie, T., Bürki, G., Pethö, L. & Michler, J. The matrix effect in TOF-SIMS analysis of two-element inorganic thin films. *J. Anal. At. Spectrom.* **35**, 1156–1166 (2020).
13. McLellan, R. A. *et al.* Chemical Profiles of the Oxides on Tantalum in State of the Art Superconducting Circuits. *Adv. Sci.* **n/a**, 2300921 (2023).
14. McLellan, R. A. *et al.* Chemical Profiles of the Oxides on Tantalum in State of the Art Superconducting Circuits. *Adv. Sci.* **10**, 2300921 (2023).
15. Ranjith, P. M., Rao, M. T., Sapra, S., Suni, I. I. & Srinivasan, R. On the Anodic Dissolution of Tantalum and Niobium in Hydrofluoric Acid. *J. Electrochem. Soc.* **165**, C258–C269 (2018).

16. Chandrasekharan, R., Park, I., Masel, R. I. & Shannon, M. A. Thermal oxidation of tantalum films at various oxidation states from 300 to 700°C. *J. Appl. Phys.* **98**, 114908 (2005).

17. Zhu, M., Zhang, Z. & Miao, W. Intense photoluminescence from amorphous tantalum oxide films. *Appl. Phys. Lett.* **89**, 021915 (2006).

18. Poorgholam-Khanjari, S. *et al.* Engineering high-Q superconducting tantalum microwave coplanar waveguide resonators for compact coherent quantum circuits. Preprint at https://doi.org/10.48550/arXiv.2412.16099 (2024).

19. Yurke, B. & Buks, E. Performance of Cavity-Parametric Amplifiers, Employing Kerr Nonlinearites, in the Presence of Two-Photon Loss. *J. Light. Technol.* **24**, 5054–5066 (2006).

20. Thomas, C. N., Withington, S., Sun, Z., Skyrme, T. & Goldie, D. J. Nonlinear effects in superconducting thin film microwave resonators. *New J. Phys.* **22**, 073028 (2020).

21. Eichler, C. & Wallraff, A. Controlling the dynamic range of a Josephson parametric amplifier. *EPJ Quantum Technol.* **1**, 1–19 (2014).

22. Anferov, A. *et al.* Low-loss Millimeter-wave Resonators with an Improved Coupling Structure. Preprint at https://doi.org/10.48550/arXiv.2311.01670 (2023).